%% file: sample-acmsmall.tex
\documentclass[acmsmall]{acmart}

\usepackage[english]{babel}
\usepackage{blindtext}
\usepackage{tikz}
\usepackage{amsmath}

\usepackage{filecontents}
\usepackage{booktabs}
\usepackage{pifont}
\usepackage{enumitem}
\usepackage{xurl}
\usepackage{multirow}
\usepackage{graphicx}
\usepackage{tabularx}
\usepackage{microtype}
\usepackage{float}
\usepackage{caption}
\usepackage{subcaption}
\usepackage{makecell}
\usepackage{wrapfig}
\definecolor{deepgreen}{rgb}{0.0, 0.5, 0.0}
\definecolor{deepred}{rgb}{0.7, 0.05, 0.0}

\newcolumntype{L}{>{\raggedright\arraybackslash}X}
\newcommand{\fcirc}[1]{\ding{\numexpr181+#1\relax}} 

\newcommand{\sys}{KVCompass\xspace}
\AtBeginDocument{%
  }

\begin{document}

\setcopyright{cc}
\setcctype{by}
\acmJournal{POMACS}
\acmYear{2026} \acmVolume{10} \acmNumber{3} \acmArticle{68}
\acmMonth{12} \acmDOI{10.1145/3857371}

\title{\sys: A Discovery Plane for Internet-Scale Prefix Cache Sharing in LLM Serving}

\author{Shiyi Ling}
\email{sling9@ucsc.edu}
\affiliation{
  \institution{University of California, Santa Cruz}
  \city{Santa Cruz}
  \state{California}
  \country{USA}
}
\author{Fei Fang}
\email{ffang6@ucsc.edu}
\affiliation{
  \institution{University of California, Santa Cruz}
  \city{Santa Cruz}
  \state{California}
  \country{USA}
}
\author{Chen Qian}
\email{cqian12@ucsc.edu}
\affiliation{
  \institution{University of California, Santa Cruz}
  \city{Santa Cruz}
  \state{California}
  \country{USA}
}


\renewcommand{\shortauthors}{Ling et al.}

\begin{abstract}
  KV cache reuse is one of the most effective ways to reduce LLM inference cost, but today its benefits are largely confined to a single cluster. When a provider serves the same model across geographically distributed regions, each region may repeatedly prefill the same popular prefixes, wasting GPU time and increasing time-to-first-token (TTFT). Exploiting reuse across regions involves discovering which worker currently holds a reusable KV cache for each request's prefix. We show that this KV cache discovery problem is structurally harder than conventional object lookup. Unlike hash-placement mechanisms such as DHTs, where an object's location is a deterministic function of its key, KV caches are created on the workers that serve requests, and their holder sets change with subsequent requests and KV cache evictions. KV cache discovery therefore calls for decoupling KV cache identity from placement. We present \sys, a discovery plane for Internet-scale KV cache sharing. The discovery index supports variable-length prefix lookup with constant-time local table reads at the edge, allowing each request to be forwarded to the worker that best balances KV cache reuse against load and network distance. Evaluation includes a 64-worker geo-distributed simulation and a 14-worker three-continent real-machine testbed. Evaluation shows that \sys achieves $1.3$--$14\times$ lower P99 TTFT than the baselines, up to $3.7\times$ lower P99 end-to-end latency, up to $4.7\times$ the throughput, and a peak load factor within $2\%$ of the Oracle floor.
\end{abstract}



\received{10 July 2026}
\received[revised]{5 October 2026}
\received[accepted]{9 October 2026}

\maketitle

\input{00_intro}
\input{01_motivation}
\input{02_design}
\input{03_implementation}
\input{04_eval}
\input{05_ablation}
\input{06_discussion}
\input{07_related}
\input{08_conclusion}

\begin{acks}
The authors were partially supported by NSF Grants 2322919, 2420632, 2426031, and 2426940. We thank our shepherd and the anonymous reviewers for their helpful feedback.
\end{acks}

\bibliographystyle{ACM-Reference-Format}
\apptocmd{\thebibliography}{\widowpenalty=10000}{}{}
\bibliography{sample-base}

\appendix
\input{09_appendix}
\end{document}

%% file: 00_intro.tex
\section{Introduction}
\label{sec:intro}

Large language models (LLMs) have rapidly become a core building block of modern computing, supporting a wide range of applications like interactive assistants, code generation, and
document analysis~\cite{vllm,316100,indatalabs2023, sglang}. As LLM adoption expands, inference workloads are growing in both scale and diversity. Processing a single request can cost orders of magnitude more than a traditional query~\cite{reuters2023, websearch-planet, tail-at-scale}, and cloud providers face increasing pressure to reduce cost while meeting strict latency requirements such as time-to-first-token (TTFT)~\cite{mooncake, preble}, making efficient LLM serving a first-order systems challenge~\cite{vllm,orca,mooncake}.

In practice, many LLM requests share common prefixes, such as system prompts, retrieved documents, or conversation histories~\cite{rag}. A large-scale measurement study at a major cloud provider~\cite{kvcache-wild} found that 10\% of KV cache blocks account for 77\% of all reuses, and that prefix-cache hit rates reach 54--62\% across representative traces. A widely adopted technique to exploit this structure to improve LLM serving efficiency is KV cache reuse~\cite{vllm,sglang,promptcache,mooncake,cachegen}: when multiple requests share a common prefix, the computed KV cache can be shared among requests, avoiding expensive prefill computation and reducing latency and serving cost. Commercial prefix caching APIs already exploit this within a single cluster, reducing cost by 50--90\% for long prompts~\cite{anthropic-cache,openai-cache}. 

However, these gains remain confined to a single cluster, while real-world LLM serving is fundamentally a global, cross-region problem~\cite{llm-deployment-wild, llm-cdn}. The demand for LLM serving is geographically dispersed: in WildChat~\cite{zhao2024wildchat}, a corpus of nearly one million ChatGPT interactions, requests span 201 countries and 178{,}724 users. At Internet scale, as major inference providers increasingly deploy the same model across tens of geographic regions to serve global users with low latency~\cite{prfaas,planetserve}, each region independently prefills the same popular prefixes, wasting GPU compute at scale. When we replay WildChat~\cite{zhao2024wildchat} in its arrival order, \autoref{fig:motivation-remoteonly} shows that $6.6\% - 21.2\%$ reusable prefixes exist \emph{only} in other regions at arrival time. However, these reuse opportunities are invisible to any KV cache confined to a single cluster or region, no matter how well it is managed. Harvesting them requires extending KV cache reuse to Internet scale, which in turn requires answering a new question for every request: \emph{which worker currently holds a reusable KV cache for this prompt's prefix?} This is the \textbf{KV cache discovery} problem, a challenge that is fundamental to Internet-scale LLM serving and unaddressed by existing systems.

KV cache discovery at Internet scale is harder than it appears. A centralized directory maps prefix hashes to KV cache locations, but each query pays a wide-area round trip that adds directly to TTFT. The natural alternative is to turn to distributed hash-based schemes such as consistent hashing~\cite{akamai} or distributed hash tables (DHTs)~\cite{chord,kademlia, pastry, can}. However, DHTs share a \textbf{fundamental limitation}: an item's placement is determined by its content hash rather than by where it was created~\cite{smash,crush}, which breaks down for KV caches: a cache entry materializes on whichever node happened to serve the prior request, not on a node designated by a hash. Forcing KV caches onto hash-designated nodes would incur migration and replication costs that negate the savings that reuse is meant to provide. KV caches thus demand \textbf{placement--identity decoupling}, which structurally rules out the entire hash-placement family, regardless of how such schemes are tuned. Therefore, no prior distributed naming or location service addresses KV cache discovery at Internet scale.

We propose a KV cache discovery plane for Internet-scale KV cache sharing. \sys is the \textbf{first system to apply space-efficient minimal perfect hashing (MPH)~\cite{mph} to Internet-scale LLM serving}. MPH's space efficiency allows the full KV cache index to be replicated in the DRAM of edge nodes in every region, making every lookup a local memory operation rather than a round trip to a remote coordinator. Its explicitly written values achieve placement--identity decoupling: the index tracks where caches land, with no control over where they go, so any worker can register a KV cache entry and any region can discover it with a local lookup. The prefix hash identifies the KV cache entry, while its value records the workers holding it and their storage tiers. Updating the value changes the holder set without changing the prefix identity. The index is organized as parallel tables at prefix lengths derived from the network latency from the edge to each region, so a lookup matches at the lengths where the prefill time a match saves exceeds the network latency of reaching its holder. Sharded maintenance units own all writes to the index, so the read path needs no locks and no cross-node coordination. At forwarding time, a joint score evaluates every worker rather than only the KV cache holders, combining network distance, queuing delay, cache tier, and residual prefill. 
Together, these properties allow \sys to reduce serving cost at Internet scale and provide the first principled answer to the KV cache discovery problem.

This paper makes the following contributions:

\textbf{(1)} We are the first to formalize the \textbf{KV cache discovery plane} as the missing layer in \textbf{Internet-scale LLM serving}, identifying its four objectives: placement--identity decoupling, geography-adaptive match granularity, local and low-overhead lookups, and load balancing, a combination that no prior distributed location service supports.

\textbf{(2)} We design and implement \sys, a prototype KV cache discovery plane built on a space-efficient hash table, with \textbf{local latency-derived multi-level prefix lookup with $O(1)$ table reads} and a \textbf{joint score} that balances KV cache reuse and worker load.
Evaluation shows \sys reduces TTFT by $1.3$--$14\times$ and end-to-end latency by up to $3.7\times$, delivers up to $4.7\times$ higher throughput, and keeps peak load factor within $2\%$ of the Oracle floor.

Section~\ref{sec:background} motivates the problem, Section~\ref{sec:design} presents the design, and Section~\ref{sec:impl} describes the implementation. Section~\ref{sec:eval} evaluates \sys. Sections~\ref{sec:discussion}, \ref{sec:related}, and \ref{sec:conclusion} discuss extensions, related work, and conclusions. Our code is available at \url{https://github.com/011shiyiLing/KVCompass}.

%% file: 01_motivation.tex
\section{Background and Motivation}
\label{sec:background}
This section motivates an Internet-scale discovery plane for cross-region KV cache reuse. We first quantify the cross-region reuse opportunity in a production workload (Section~\ref{sec:background:crossregion}), then derive four requirements for an effective discovery plane (Section~\ref{sec:background:objectives}), and finally explain why directly deriving cache locations through hash-based placement does not satisfy these requirements (Section~\ref{sec:background:hash}).
\subsection{Cross-region LLM Serving}
\label{sec:background:crossregion}
An LLM serving system~\cite{vllm,sglang} hosts model replicas on multiple GPU workers and responds to each user prompt by generating a stream of output tokens.
A request is processed in two phases: a \emph{prefill} phase that receives the prompt and computes its \emph{KV cache}, followed by a \emph{decode} phase that generates one token per step. Serving quality is measured along both the latency and throughput dimensions. Users experience time to first token (TTFT), which is largely determined by queueing and prefill, as well as end-to-end latency, which additionally includes autoregressive decoding. Providers care about the aggregate throughput sustained by the worker pool, as it directly affects the serving cost per request~\cite{vllm,distserve}.

\begin{wrapfigure}[10]{r}{0.5\textwidth}
  \centering
  \includegraphics[width=\linewidth]{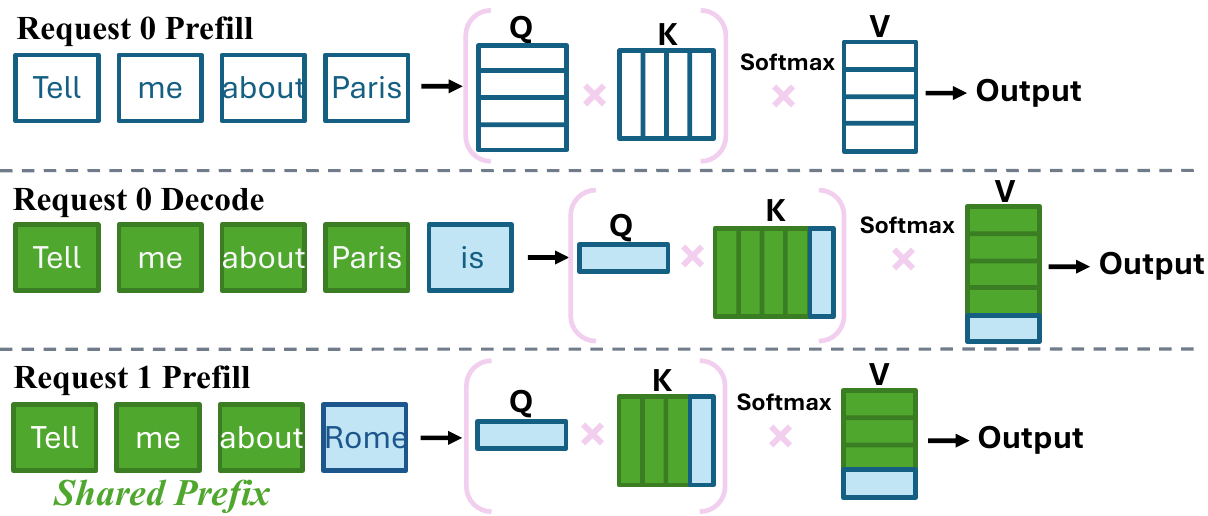}
  \caption{KV cache reuse via prefix sharing.}
  \label{fig:prefix-caching}
  \Description{Diagram of two requests sharing a prompt prefix. The second request reuses the KV cache of the shared tokens and prefills the remaining tokens.}
\end{wrapfigure}

Prompts frequently share long common prefixes, such as system prompts, few-shot templates, and multi-turn history. 
Therefore, modern engines retain the KV cache of served prefixes and skip prefill for the shared part, simultaneously improving TTFT and throughput~\cite{sglang,mooncake,anthropic-cache,openai-cache}. As illustrated in \autoref{fig:prefix-caching}, request~0 first prefills its prompt and uses the resulting KV cache for decoding. Because identical prefixes produce identical KV states, request~1 can reuse the KV states of the three shared tokens and prefill only the remaining tokens. Longer shared prefixes yield greater prefill savings.

However, these gains are confined to a single cluster within one region that holds the KV cache. Nowadays, production providers deploy serving clusters in geo-distributed regions. These clusters serve the same model independently, and each keeps its own local KV cache. The natural next step is to extend KV cache reuse to global scale.
This raises a question: \emph{do user requests in one region share prefixes already cached in another?}
\begin{figure*}[t]
  \centering
  \begin{minipage}[t]{0.49\linewidth}
    \centering
    \includegraphics[width=\linewidth]{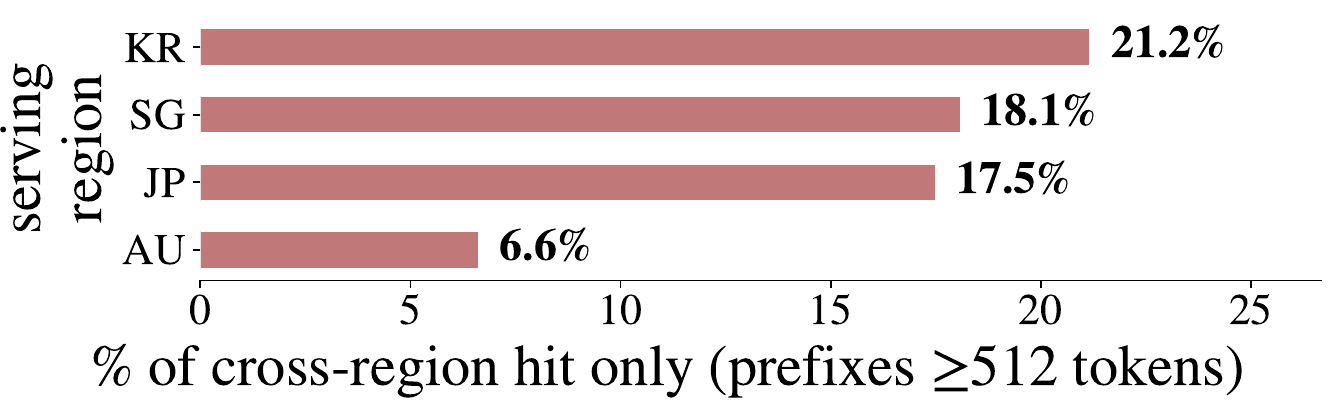}
    \captionof{figure}{Cross-region only KV cache hit of ${\geq}512$-token prefixes in WildChat~\cite{zhao2024wildchat}}
    \label{fig:motivation-remoteonly}
    \Description{Bar chart of reusable WildChat prefixes held only in other regions at request arrival, showing cross-region KV cache reuse opportunities.}
  \end{minipage}
  \hfill
  \begin{minipage}[t]{0.49\linewidth}
    \centering
    \includegraphics[width=\linewidth]{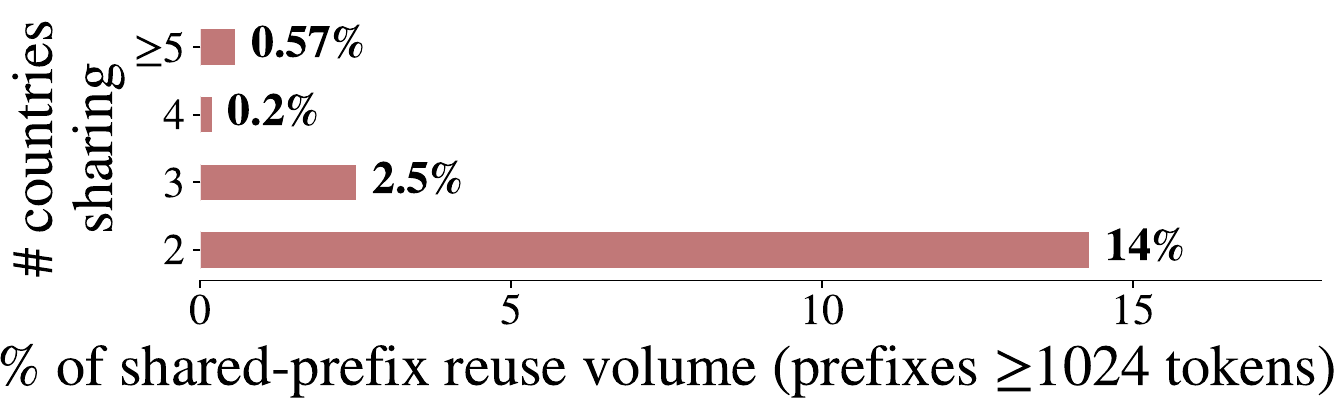}
    \captionof{figure}{Cross-region reuse in WildChat~\cite{zhao2024wildchat} for prefixes of ${\geq}1024$ tokens.}
    \label{fig:motivation-cross}
    \Description{Bar chart of shared-prefix reuse volume in WildChat grouped by the number of countries sharing a prefix.}
  \end{minipage}
\end{figure*}

To answer this question, we analyze the WildChat~\cite{zhao2024wildchat} workload.
We take the first 1,024 tokens of each request as its prefix.
We keep the prefixes reused by at least two distinct users, and map each request's origin country to the nearest of 8 deployment regions.
A shared prefix is \emph{cross-region} when its sharing users map to at least two different regions; only these reuses face a choice between recomputing the prefix locally and reusing a KV cache in a remote region. \autoref{fig:motivation-cross} shows the volume of the cross-region prefixes, bucketed by how many countries share the prefix. Prefixes shared between two countries carry 14\%, while those shared between three countries carry 2.5\%. Overall, approximately 17\% of all shared-prefix reuse volume crosses a single region boundary. 

However, in today's deployments a router only knows the KV caches stored on workers in its own region, so it cannot find a remotely computed KV cache and has to recompute the prefix from scratch. This 17\% of the potential reuse thus has no chance to happen.
Losing this share also costs the most, since prefixes shared across regions are usually the long, popular ones whose recomputation is the most expensive. \textbf{Cross-region KV cache reuse} is therefore both large in volume and worth exploiting per request.
Thus, the \emph{KV cache discovery at Internet scale}, looking up which worker anywhere in the world already holds a reusable KV cache, becomes an important challenge. Appendix~\ref{app:traces} provides additional reuse analysis on Qwen-Bailian~\cite{kvcache-wild} and Mooncake~\cite{mooncake} production traces.

\subsection{Objectives for a KV Cache Discovery Plane}
\label{sec:background:objectives}
The preceding analysis establishes a substantial opportunity for cross-region prefix KV cache reuse. Realizing this opportunity, however, requires a discovery plane that can locate reusable caches across regions without adding significant latency or creating new load imbalances.

Suppose a provider operates a serving worker pool across $R$ geo-distributed regions. The KV cache for a prefix may be stored in a different region's worker from the request that could reuse it.
Realizing cross-region reuse therefore requires an \textbf{Internet-scale KV cache discovery plane}. For each incoming request, this layer identifies workers that currently hold reusable prefix KV caches and combines those matches with network and load information to select the locally estimated fastest serving worker. To discover reusable KV caches across regions correctly and efficiently, \sys needs to satisfy four objectives.

\noindent\textbf{(O1)~KV caches need to be discoverable wherever the requests that compute them land, not where their identities hash to.} In conventional location services using hash-based placement, an item lives where its identity dictates: a content hash maps the item to a designated node, and a lookup recomputes that mapping to find it.
However, a KV cache entry obeys no such rule. It is not a deliberately placed object, but a by-product of computation that materializes on whichever worker happened to serve the request that produced it.
Therefore, a KV cache discovery plane cannot derive a KV cache's location from its prefix's hash value; it needs to locate KV caches wherever they have been served, with no control over where they go. We call this requirement \emph{placement--identity decoupling}. 

\begin{figure*}[t]
  \centering
  \begin{minipage}[t]{0.50\linewidth}
    \centering
    \includegraphics[width=\linewidth]{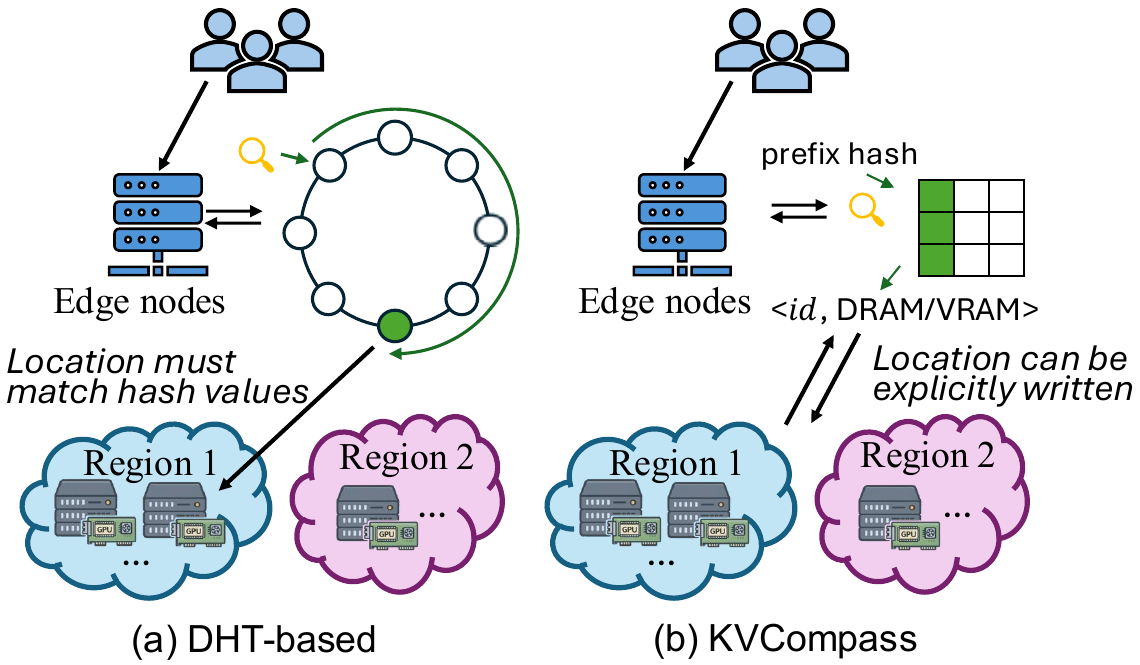}
    \captionof{figure}{Comparison of hash-placement and explicit-write
    approaches for KV cache discovery.}
    \label{fig:motivation}
    \Description{Diagrams comparing hash-directed KV cache placement with an explicit-write discovery index that records holders wherever KV caches are created.}
  \end{minipage}
  \hfill
  \begin{minipage}[t]{0.48\linewidth}
    \centering
    \includegraphics[width=\linewidth]{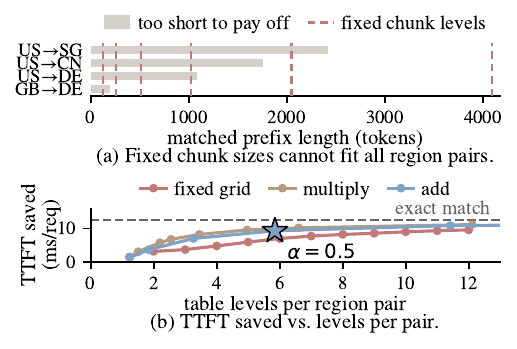}
    \captionof{figure}{Network latency-derived index levels versus a fixed
    chunk grid.}
    \label{fig:levels}
    \Description{Plots comparing latency-derived prefix lengths with fixed grids and showing the reuse benefit captured as the number of indexed levels increases.}
  \end{minipage}
\end{figure*}
\noindent\textbf{(O2)~Prefix match lengths need to adapt to both the request content and geography.} A prompt of a request does not share a fixed number of tokens with the KV caches stored on different workers.
For example, a 6000-token prompt may share its first 256 tokens with some cached entries, its first 2048 with others, and its first 5120 with a third set~\cite{parrot}.
Therefore, looking up every candidate at one fixed length considers only a fraction of the matches longer than that length, and misses the matches shorter than the length entirely. Extending the index to a fixed length of multiple chunk sizes does not resolve this problem either, because at Internet scale the length worth matching is not a property of the prompt itself alone. Reusing a KV cache in a remote region pays off only when the prefill time it saves exceeds the network latency of reaching the worker holding the KV cache, and that latency depends on which region the request is forwarded to. As \autoref{fig:levels}(a) shows, the shortest prefix length worth forwarding ranges from a few hundred to a few thousand tokens, depending on the distance between regions. Suppose every region used one fixed grid of lengths. For any given region,
many of the grid's lengths then fall below its shortest worthwhile
length toward some remote region, and a match found at those
lengths saves less prefill time than the forwarding latency it
pays, so these are matches we do not want to discover.
Thus, the discovery plane needs to judge each candidate at match lengths that reflect both the prompt's sharing structure and the geographic distance between the user and the KV cache holder.

\noindent\textbf{(O3)~Lookups need to be local and effectively free.} A lookup that triggers a network round trip to a remote node adds directly to TTFT, eliminating the time savings that KV cache reuse is meant to provide. Thus, every lookup has to resolve locally at the edge node, without off-node coordination. At Internet scale this imposes two demands. First, the state behind a lookup needs to be small enough to reside at every edge node rather than at a centralized directory.
Second, each lookup needs to be fast. Because an edge node performs a lookup for every request at tens of candidate prefix lengths, the index answers tens of queries for every request served. Multiplied across all requests, even a small cost per lookup becomes the dominant work on the edge, and an overloaded edge stalls every request behind it. 
O2 sharpens this objective further. Because the worthwhile match length varies with the geographic location of each holder, a lookup cannot settle for a single fixed depth, yet it has to check all relevant lengths within this same tight local budget.

\noindent\textbf{(O4)~Discovery needs to avoid creating hotspots.} Popular prefixes are the most valuable to discover, and they also attract the most requests. But forwarding every matching request to the few workers that already hold a popular prefix's KV cache will overload those workers while the rest of the workers sit idle. As requests concentrate on one worker, its queueing delay grows and eventually outweighs the prefill time saved by reusing the KV cache. Therefore, the discovery plane needs to weigh this saving against each worker's load and network distance, and may prefer an idle worker that recomputes the prefix over a congested or distant one.
\begin{figure*}[t]
  \centering
  \begin{minipage}[t]{0.42\linewidth}
     \centering
    \includegraphics[width=\linewidth]{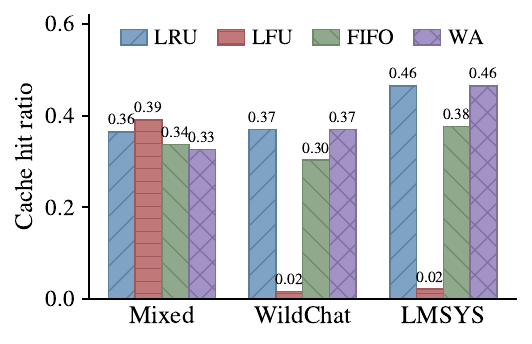}
    \captionof{figure}{KV cache hit ratios by eviction policy.}
    \label{fig:evict}
    \Description{Grouped bars comparing KV cache hit ratios under four eviction policies on Mixed, WildChat, and LMSYS. The best policy differs by workload.}
  \end{minipage}
  \hfill
  \begin{minipage}[t]{0.57\linewidth}
    \centering
      \includegraphics[width=\linewidth]{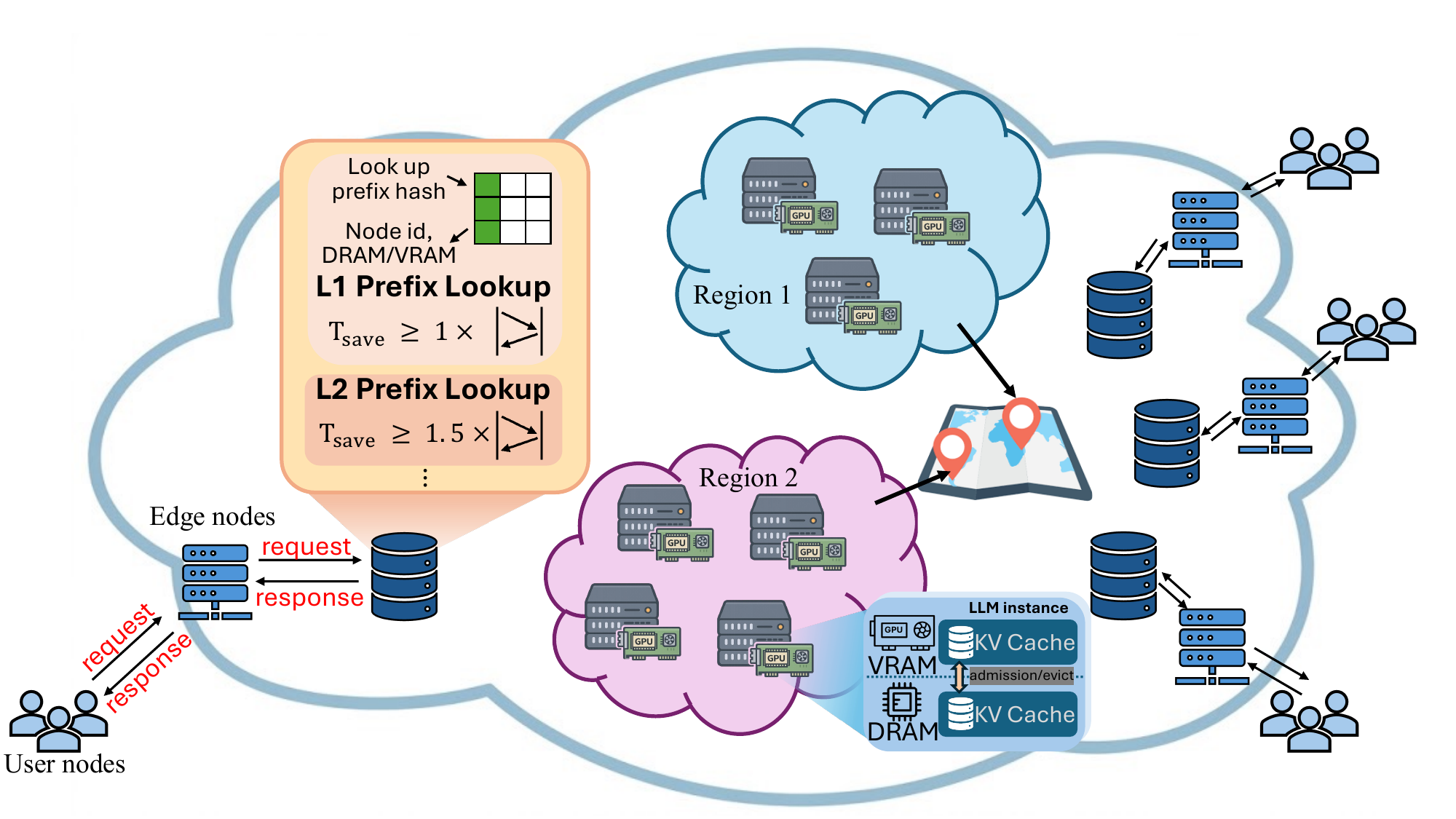}
      \caption{\sys overview.}
      \label{fig:overview}
      \Description{System diagram showing edge lookup units, maintenance units, and workers across regions. Requests use local discovery index views, and workers report KV cache events to update those views.}
  \end{minipage}
\end{figure*}

\subsection{Why Hash-Directed Cache Placement Is Insufficient}
\label{sec:background:hash}

A natural idea for a location service in an Internet-scale system is a DHT such as Kademlia~\cite{kademlia} or Chord~\cite{chord}, a distributed implementation of consistent hashing used within clusters for prefix KV cache affinity~\cite{mooncake}.
However, they all assume \emph{placement equals identity}: an item's location is determined by its hash value, as Figure~\ref{fig:motivation}~(a) illustrates, the assumption that O1 rules out.
To mediate a KV cache's created and hash-designated locations, a hash-placement scheme would have to migrate each entry to its hash-designated node, which costs hundreds of megabytes of wide-area bandwidth per entry.
In the other direction, the scheme would index only the entries that already sit at their hash-designated nodes; since KV caches are rarely at these nodes, most KV caches would become undiscoverable.
Therefore, the entire hash-placement family, including Chord, Kademlia, consistent hashing, and Rendezvous hashing, is invalid as a foundation for this discovery plane.

The KV cache discovery plane instead needs a data structure whose value field is \emph{explicitly written} by whoever holds the KV cache, not derived from the key, as Figure~\ref{fig:motivation}~(b) illustrates.
Hashing may decide where a mapping is stored \emph{inside the index}, but never where the KV cache itself lives.
Such an explicit-write structure satisfies placement--identity decoupling~(O1) by construction.
Section~\ref{sec:design} shows how \sys builds on this principle to meet the remaining objectives.


%% file: 02_design.tex
\section{Design of \sys}
\label{sec:design}
\subsection{System Overview}
\label{sec:design:overview}
We target an LLM serving deployment that spans \emph{multiple regions}.
Regions are geographically distributed, in different countries or even continents.
Each region hosts a set of \emph{workers}.
A worker is one inference instance, holding the full model replica on one or more GPUs, and it keeps the KV caches of the prefixes it has served in its VRAM and host DRAM.
A worker reports a \emph{KV cache event} whenever its local KV cache changes state, admitting a newly computed prefix, moving one between VRAM and DRAM, or evicting one.
Workers within a region share a high-speed datacenter network, while traffic between regions crosses the wide-area network, with round-trip times orders of magnitude larger.
Each region also runs several \emph{edge nodes}, CPU-only gateways separate from the workers.
A user request arrives at its nearest edge node, and is forwarded to a worker, in the same region or in a remote region.

\sys is a KV cache discovery plane for this distributed LLM serving.
Forwarding every request to the few workers holding popular prefixes would create the hotspots that~(O4) warns against, so \sys balances KV cache affinity and worker load jointly. For each incoming request, it identifies which workers across all regions hold a reusable KV cache for the request's prefix, then weighs the saved prefill cost against each candidate's load and geographic network latency, and forwards the request to the worker with the lowest expected completion time.

\sys is built from a single logical data structure and two kinds of software components that act on it.
The data structure is the \emph{multi-level ludo discovery index} (Section~\ref{sec:design:index}), which maps prefix hashes to the workers currently holding the corresponding KV cache.
\emph{Maintenance units} are the sole writers of the index and apply KV cache events to it.
\emph{Lookup units} answer the discovery query of every request: which workers hold a reusable KV cache for its prefix, and in which storage tier; each holds a local view of the discovery index, a compact slot-only copy without the keys (Section~\ref{sec:design:components}).
The components act on the discovery index through four operations, \texttt{lookup} for forwarding the request and \texttt{insert}, \texttt{update}, and \texttt{remove} for the KV cache events (Section~\ref{sec:design:ops}).
We also analyze the memory footprint, operation costs, and control traffic of \sys in Section~\ref{sec:design:perf}.
Table~\ref{tab:notation} summarizes these terms and the notations used in the rest of the paper.

Figure~\ref{fig:overview} shows the request and update flows. At the nearest edge node, the lookup unit queries latency-derived prefix lengths and returns matching workers and their KV cache tiers. \sys scores all workers across regions using residual prefill time, load, and network distance, then forwards the request to the winner. Workers report KV cache admissions, tier changes, and evictions to maintenance units, which update the discovery index and lookup units.

\sys does not control when a worker admits, demotes, or evicts a KV cache. \autoref{fig:evict} compares four eviction policies on three workloads. LFU achieves the highest KV cache hit ratio on Mixed, while LRU performs best on WildChat and LMSYS. With no policy best across workloads, \sys leaves the choice to each worker and only tracks the resulting KV cache events.

\begin{table}[t]
  \centering
  \small
  \caption{Terms and notations.}
  \label{tab:notation}
  \begin{tabular}{@{}c>{\centering\arraybackslash}p{0.46\linewidth}c@{}}
    \hline
    \textbf{Term / Notation} & \textbf{Meaning} & \textbf{Example values} \\
    \hline
    worker           & inference instance holding KV caches & $\sim 10$--$10^2$ \\
    \hline
    region           & geographic site hosting a set of workers  & $\sim 10$ \\
    \hline
    edge node        & gateway of a region, hosting a lookup unit & several per region \\
    \hline
    lookup unit      & keyless view of the discovery index & one per edge node \\
    \hline
    shard            & a logical group of the discovery index entries, assigned by hashing the key; & $10^{4}$--$10^{6}$ entries per shard \\
    \hline
    maintenance unit &  applies KV cache events to its shard and pushes the updates to the lookup units & one per shard \\
    \hline
    $L$              & the number of the discovery index levels  & ${\sim}6$ per edge node \\
    \hline
    $E$              & the total number of the discovery index entries & $10^{5}$--$10^{7}$ \\
    \hline
    $n_1$            & first latency-derived prefix length & $10^2$--$10^3$ tokens \\
    \hline
    $\mathrm{WAN}(o,r)$ & one-way latency from region $o$ to worker $r$'s region & ${\sim}10^2$\,ms \\
    \hline
    $\beta_r$        & worker $r$'s prefill rate  & ${\sim}0.1$\,ms/token \\
    \hline
  \end{tabular}
\end{table}

\subsection{The Multi-Level Ludo Discovery Index}
\label{sec:design:index}
The discovery index is the shared state of \sys: lookup units read it and maintenance units write it. We first recall the hash structure it builds on and why it is suitable for our problem, then describe the modifications that the KV cache discovery plane demands of the discovery index.

The index builds on Ludo hashing~\cite{ludo}, a key-value lookup engine that combines a minimal perfect hash function (MPH)~\cite{mph} with cuckoo displacement: given any set of key-value mappings, it constructs a compact lookup structure that returns a key's value in $O(1)$ with a few memory accesses, without storing the keys themselves, cutting space by up to 90\% compared with state-of-the-art hash tables. The Ludo table is sized at build time for its expected entry count and does not grow in place. 
Its use of hashing is fundamentally different from the hash-placement family dismissed in Section~\ref{sec:background:hash}: Ludo hashes a key only to locate the table slot holding a mapping that the writer chose, so the value can name any node, and two workers that independently cache the same prefix register their own locations under the same key, with no migration and no structural conflict. The explicitly written value yields the placement--identity decoupling that~(O1) calls for, letting the plane track KV caches wherever computation leaves them. The keyless compactness makes parallel tables at many match lengths affordable, so multi-granularity matching~(O2) need not multiply memory cost, and, combined with $O(1)$ lookup, lets the discovery index replicate in the local memory of every edge node, keeping lookups local and effectively free~(O3).

However, a flat Ludo table does not yet satisfy these objectives. As a key-value design it maps each key to a single value, hence exactly one worker, while in Internet-scale sharing the same prefix is often cached by many workers at once. It operates at one fixed prefix length, treating a nearby holder and a distant one equally, and because it stores no keys, a lookup for a prefix that was never inserted can return a false hit. Therefore, \sys redesigns the structure in three ways, which are described below.

\noindent\textbf{Tracking dynamic, multi-holder KV cache sets.} The same prefix is cached by whichever workers happened to serve requests for it; the holder set churns continuously as KV caches are admitted under load and evicted under memory pressure, and each holder independently moves the prefix between VRAM and host DRAM, so a discovery index needs to resolve to a mutable set of holders and their storage tiers under a high rate of updates. \sys organizes the index as Ludo tables. Each entry indexes one prefix at one length within its table. For example, the entry indexing the first $c$ tokens of a prompt is keyed by a 63-bit key, the truncated SHA-256 digest of those $c$ tokens; $c$ here is drawn from the network latency-derived set of indexed lengths constructed below, so entries at many different lengths coexist in the same table. Each entry stores a packed value with one 2-bit tier field per worker in its table. The field is \texttt{01} for \textsc{vram}, \texttt{10} for \textsc{dram}, and \texttt{00} otherwise. Each region has its own table. A 32-bit value holds tier fields for up to 16 workers, and a 64-bit value holds up to 32. Each field uses 2 bits. Every edge node keeps all region tables and combines their lookup results to find the holders. A region with more than 32 workers requires additional tables. When an entry is created, all worker fields in its table are initialized to \texttt{00}. A worker's field becomes \texttt{01} or \texttt{10} when it caches the prefix. One read returns all worker fields in that table, and a tier change is an in-place 2-bit write. We exclude SSD because fetching a prefix on request arrival adds an SSD read to the serving path, typically costing more than recomputing it.

\noindent\textbf{Matching prefixes at distance-aware granularities.} The discovery index needs to answer one question after finding the matched prefix: is this prefix match long enough to be worth forwarding to that region to reuse? Fixed chunk sizes of prefix cannot answer it~(O2). Forwarding to another region to serve the request adds the round trip from the edge to the target region, while the match buys skipped prefill time, which grows linearly with the matched tokens at the per-worker prefill rate $\beta$ in ms/token. A match of $n$ tokens saves about $\beta n$ ms, so this forwarding decision pays off once this KV cache reuse saving covers the added round trip. Therefore, \sys cuts the first indexed length at exactly this point. This shortest profitable prefix length is the first \emph{latency-derived prefix length} $n_1$, and it scales with the measured latency from the edge to that region. Within the edge's own region the round trip is so small that this formula would put $n_1$ at a few dozen tokens, guarding a cost that hardly exists, so same-region matching starts at a short fixed minimum.

But one level per region is still not enough. The first latency-derived prefix length only tells the edge whether that forwarding is worthwhile; to pick among candidates, the forwarding decision needs to know how much each match saves. With a single indexed length a very long token match and a barely profitable one would look identical. \sys thus indexes several levels of increasing lengths above each region's $n_1$, with the same number of levels toward every region. \autoref{fig:levels}(b) compares different choices of the level lengths over the WildChat~\cite{zhao2024wildchat} reuse distribution. Latency-derived levels beat latency-blind fixed grids at every table budget, and lengths spaced at equal steps of half of $n_1$, giving $n_1$, $1.5\,n_1$, $2\,n_1$, and so on, capture about three quarters of the exact-match upper bound with about six levels. \sys adopts these steps. We also measured multiplicative growth, but a match falls to the nearest indexed length below it, and multiplicative spacing grows coarsest exactly where matches are longest and most valuable, whereas the equal-step spacing keeps the under-crediting bounded by half of $n_1$ at every match length. The same-region lengths instead double from the short minimum, following the chunk sizes by which workers organize their KV caches. All the level lengths derive from a stable WAN latency matrix and per-worker prefill rates, so the whole structure can be computed offline and adds no extra overhead to the request serving path.

\noindent\textbf{Suppressing false hits without inflating the discovery index size.} Because Ludo does not store the keys it indexes, a lookup for a prefix that was never inserted may collide with an unrelated entry and name a worker that does not actually hold it, so the request is forwarded to a worker without the prefix and pays a needless WAN round trip. Storing full keys at every edge node would eliminate these false hits but inflate the discovery index far beyond DRAM budgets, defeating the space efficiency that makes edge local lookup possible. \sys instead lets a false hit surface and absorbs it cheaply: the worker a request is forwarded to serves only the prefix it truly holds and recomputes the rest, so the answer is always correct and the only cost is one forwarding that saves less prefill than expected. To keep such forwardings rare, each edge node discounts a worker's claimed match by a \emph{cache belief} maintained from past responses, so a worker that often does not actually hold what it claims stops attracting requests, while the maintenance units keep the full keys off the forwarding path and use them only for collision detection when updating the index.

\subsection{System Components}
\label{sec:design:components}
There are two kinds of software components that act on the discovery index. Lookup units resolve which workers hold how much of a prompt's prefix at the edge node. Maintenance units turn KV cache events into the discovery index writes and propagate them to the lookup units.

\noindent\textbf{Lookup units.}
A lookup unit runs on the edge node and holds a view of every shard. Each lookup unit queries the multi-level tables at multiple lengths, derived from its measured network latencies toward each region. The lookup view stored on every edge node is identical. A worker registers each cached prefix at the union of lengths that any edge may query, so only the queried lengths differ across regions. Identical views keep the edges interchangeable. A new or recovering edge can bootstrap from the maintenance units or copy any other edge, and if every edge node of a region fails, another region's edge can stand in. 
Holding everything at every edge node also costs little, since a view does not store the keys. Answering a query needs only the slot states, so the keys stay with the maintenance units. 

When a request arrives, \sys computes the prefix digests at the edge's query lengths, reads the $\langle\mathit{worker},\mathit{tier}\rangle$ holders at each length from the lookup unit's view, evaluates the joint score over all workers across regions, and forwards the request to the winner. Queries are plain memory reads with no locking, because each shard has a single writer and its update messages are applied in the order they were sent.

\noindent\textbf{Maintenance units.}
Maintenance units construct and update the discovery index and are its sole writers. Each unit owns one shard, the keys whose $\mathrm{hash}(K)\bmod N$ matches its identifier, and runs as a CPU-resident process co-located with a worker. Shards split the key space among the $N$ writers, while views replicate the whole key space at every edge. Each maintenance unit therefore sends its shard's updates to every edge when KV cache events happen, and each edge receives the update information of all $N$ units. 
On each KV cache event, the unit locates the key in its shard, applies the implied 2-bit field edit, and sends the update to the lookup unit, so the lookup views stay in sync without communicating with each other. 

\begin{figure}[t]
  \centering
  \includegraphics[width=\linewidth]{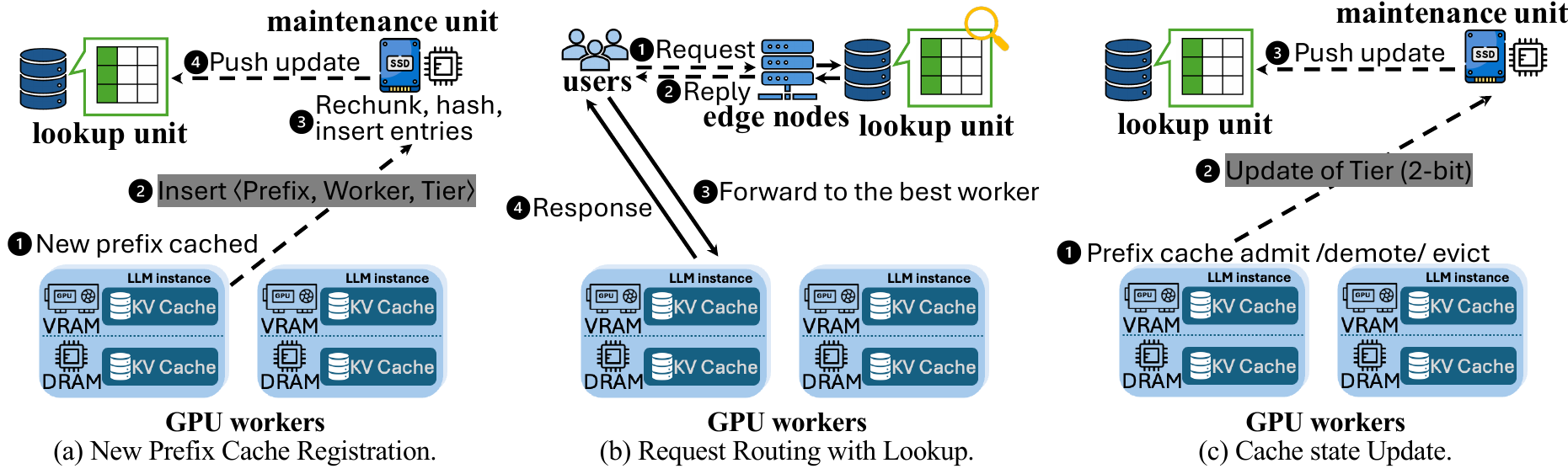}
  \caption{Discovery plane operations. (a)~A worker registers a newly cached prefix. (b)~The edge node looks up a request and forwards it to the best worker under its local view, as determined by the joint score. (c)~A KV cache state change updates the worker's 2-bit storage-tier field.}
  \label{fig:operation}
  \Description{Three operation diagrams showing prefix registration, request lookup and forwarding, and KV cache tier updates propagated from maintenance units to edge lookup units.}
\end{figure}

\subsection{Operations of the Discovery Index}
\label{sec:design:ops}
\sys exposes four operations on the discovery index: \texttt{lookup} when a user sends an inference request at an edge node; \texttt{insert}, \texttt{update}, and \texttt{remove} when a worker's KV cache changes state. \autoref{fig:operation} illustrates the flow of each, and we describe them in turn.

\noindent\textbf{\texttt{Lookup}.}
As \autoref{fig:operation}(b) shows, a user request lands at the edge node closest to its origin region $o$~\fcirc{1}, the lookup unit answers with the matching workers and their tiers~\fcirc{2}, and the request is forwarded to the best worker~\fcirc{3}, which will return the response to the user ~\fcirc{4}. To find the matches, \sys scans the discovery indexes on the Ludo table at every level $l$ whose indexed length $c_l$ fits the prompt, obtaining for each worker $r$ its longest matching prefix length $m_r$ and storage tier $t_r$. \sys ranks candidates with a joint score that weighs each worker's load, KV cache affinity, and network and reload cost in one formula. The score is evaluated over all workers across regions rather than only the KV cache holders. Under load, an idle worker that recomputes the prefix from scratch frequently beats a distant or congested holder, the preference that~(O4) calls for. The discovery plane forwards the prompt to the worker minimizing $\mathrm{score}(r) = \mathrm{WAN}(o,r) + \widehat{T}_{q}(r) + (1+\lambda)\beta_r (L_{\mathrm{prompt}} - b_r m_r) + \gamma\, b_r m_r\, \mathbf{1}[t_r = \textsc{dram}]$,
where $\widehat{T}_{q}(r)$ is the estimated queueing delay of $r$,
$\beta_r$ is the per-worker prefill rate in ms/token, and $\gamma$ is the per-token PCIe reload cost, charged only when the matched prefix is DRAM-resident. Worker $r$'s cache belief $b_r \in [0,1]$ is the exponentially weighted moving average of the ratio of actually reused prefix tokens, reported in each response, to the match claimed by the edge. Frequent false claims lower $b_r$ and the effective match $b_r m_r$, making the worker less likely to win the joint score.

The score consults $\widehat{T}_{q}(r)$ for workers in every region, yet an edge observes only its own region directly. Each edge node therefore keeps a per-worker queue estimate, read live for the local region and refreshed for remote regions through two channels. A cross-region response carries the serving region's per-worker queue vector, which the requesting edge folds into its estimate with an exponentially weighted moving average of weight $w$. Each region also broadcasts its vector at a fixed period $\Delta$, and the broadcast is suppressed toward any region that recently received a piggybacked report. Staleness is thus bounded on both paths, within one response interarrival for a busy pair and one broadcast period
for an idle one. Freshness also matches demand, since the pairs whose
forwarding depends most on accurate estimates are exactly the pairs
refreshed most often. The queue vector adds only a few dozen bytes to responses already crossing the WAN, so the fresher estimates cost no additional control traffic.

The factor $(1+\lambda)$ prices recomputation by spare capacity. With $\lambda = \min(\lambda_{\max}, \lambda_{q})$, where
$\lambda_{q} = \min_{r'} \widehat{T}_{q}(r') / \tau$ for a normalization constant $\tau$, the capacity price stays near zero while idle workers exist and grows as the pool saturates. On an idle worker, recomputation costs only the request's own prefill time and often beats a WAN round trip to a distant holder. At saturation every recomputed token delays queued requests, so reuse saves aggregate work and the inflated multiplier shifts the choice toward KV cache holders. Each edge computes $\lambda$ from its own estimates, so the pricing reflects the spare capacity that particular edge can see when it forwards. A stale estimate can misprice $\lambda$, but the error is bounded, since $\lambda$ is capped at $\lambda_{\max}$ and $1/\tau$-Lipschitz in the estimates, and it skews at most that one forwarding decision.

\noindent\textbf{\texttt{Insert}.}
As \autoref{fig:operation}(a) shows, when a worker completes a new prefill, the prompt's prefix is now in its VRAM~\fcirc{1}, and the worker registers this new copy with the discovery plane. The worker reports one registration event carrying the prefix, its own identifier, and the tier~\fcirc{2}. The maintenance units rechunk the prefix at every indexed length $c_l \le L_{\mathrm{prompt}}$ and hash each chunk into its key $K_l$, and the maintenance unit owning $K_l$ inserts the entry by rewriting the worker's 2-bit slot to \texttt{01}~\fcirc{3}. The corresponding update message is propagated to the lookup unit~\fcirc{4}: a short cuckoo-move path if the key was previously absent, or a single in-place slot patch otherwise.

Request serving is not the only trigger of inserts. Once a worker serves a prefix, the resulting \texttt{insert} makes that copy discoverable, so popular prefixes spread across the worker pool as a byproduct of serving. However, this passive spread fails when a prefix turns hot in a region that has never served it, where every early request must either cross the WAN to a remote holder or recompute from scratch. \sys covers this transient by migrating a popular KV cache replica proactively, and the migration triggers an \texttt{insert} from the destination worker. \sys tracks the popularity of prefixes locally at each edge node, counting per \emph{prefix family}, all the prefixes sharing the same first-chunk key. A prefix family turns hot once its local misses reach $H$ within a sliding time window $T_h$. What migrates is the triggering request's own prefix, which contains the shared ancestor, so every sibling in the same prefix family benefits at least up to the shared portion. The migration is triggered when the region has no local holder and a remote holder's match exceeds the latency-derived prefix length of that holder's region, or when every local holder keeps losing the joint score to load. To stay off the request serving path, the replica goes only to a persistently idle worker, one whose estimated queue has stayed at zero, and it lands in the destination worker's DRAM so as not to evict the KV caches already serving from VRAM. Within a region the replica is copied worker-to-worker connected by a high-bandwidth network. Across regions moving KV tensors over the WAN is uneconomical, so \sys ships the prompt and recomputes the KV cache at the destination worker, and since the destination worker is idle, the recomputation takes no GPU time from serving. 

\noindent\textbf{\texttt{Update}.}
As \autoref{fig:operation}(c) shows, when a worker demotes a prefix from VRAM to host DRAM or promotes it back, it reports a KV cache event for the prefix~\fcirc{1}. The owning maintenance units rewrite that worker's 2-bit field at every level's discovery index~\fcirc{2} and propagate the update to the lookup unit~\fcirc{3}. The prefix is already resident from its earlier insertion, but a cuckoo displacement may have moved it since, so the unit searches both candidate buckets for the current slot before writing the new tier code. No cuckoo move is involved, because an update changes only the stored value and relocation runs only when an insertion hits a full bucket. A single resulting
$(\mathrm{bucket},\mathrm{slot},\mathrm{value})$ triple is propagated to the lookup unit. 

\noindent\textbf{\texttt{Remove}.}
When a worker fully evicts a prefix, the maintenance unit clears its slot to \texttt{00} along the same path as in \autoref{fig:operation}(c)~\fcirc{2}. If other workers still hold the prefix, the entry remains; once the packed value drops to zero, the entry is removed from the table. No tombstone is required: each shard has a single writer, and stale retransmissions are rejected by the per-rank sequence check.

\begin{figure}[t]
  \begin{minipage}[t]{0.36\textwidth}
    \centering
    \includegraphics[width=\linewidth]{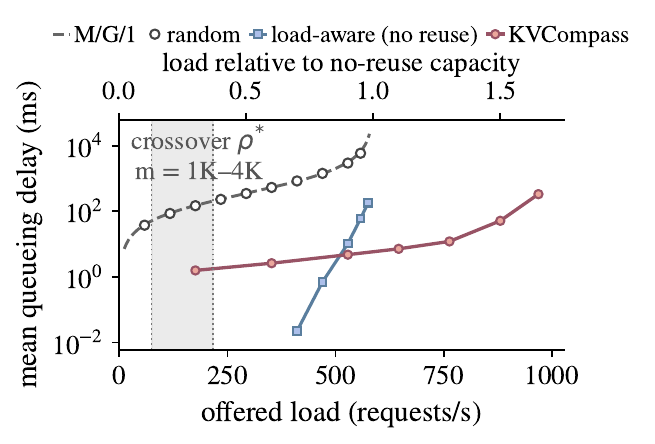}
    \caption{Simulated mean queueing delay versus offered load (Mixed).}
    \label{fig:mg1}
    \Description{Plot of simulated mean queueing delay versus offered load on Mixed, comparing the queueing model and routing policies with and without KV cache reuse.}
  \end{minipage}
  \hfill
  \begin{minipage}[t]{0.63\textwidth}
    \centering
    \includegraphics[width=\linewidth]{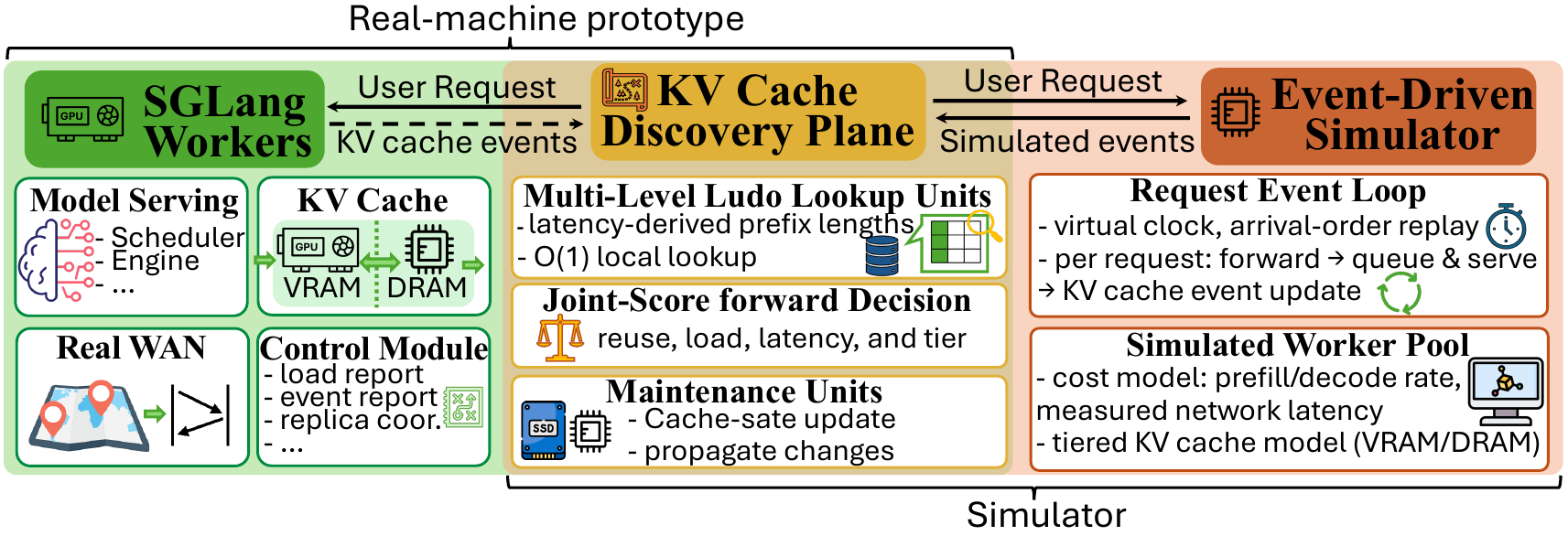}
    \caption{\sys implementation. The prototype and event-driven simulator share the same KV cache discovery plane.}
    \label{fig:implementation}
    \Description{Architecture diagram showing the real-machine prototype and event-driven simulator sharing discovery-plane components while using different worker and network execution paths.}
  \end{minipage}
\end{figure}

\subsection{Performance Modeling and Analysis}
\label{sec:design:perf}
We analyze discovery index memory, operation costs, and control traffic. $E$ is the total entry count across all tables, $W$ the worker count, $R$ the region count, $L$ the number of indexed levels, and $\beta$ the prefill time per token in ms.

\noindent\textbf{Memory footprint.}
Following the Ludo analysis~\cite{ludo}, at 95\% occupancy a table requires about 0.263 four-slot buckets and 2.33 classifier bits per entry. With 32-bit values, the lookup view costs $2.33 + 0.263 \times (5 + 4 \times 32) \approx 37$ bits per entry. Retaining 63-bit keys raises the maintenance cost to about 104 bits per entry. With 64-bit values, these costs rise to about 71 and 137 bits per entry. Total memory sums these costs over all table entries, with additional allocation overhead. As workers and regions are added, more tables may be needed, and prefixes held in multiple tables occupy separate entries. Each edge stores all tables, so more edges increase total lookup-view memory.

\noindent\textbf{Lookup cost.}
A lookup makes at most $L$ table reads per region, one for each indexed length that fits the prompt. Each read takes about three memory accesses, two for the classifier and one for the bucket, independent of $E$. Across $R$ regions, table reads cost $O(LR)$ and scoring costs $O(W)$, giving $O(LR + W)$ per request. All reads are local; only request forwarding crosses the WAN.

\noindent\textbf{Queueing delay under load.}
A queueing approximation explains when load outweighs KV cache reuse in the joint score. Model a worker as an M/G/1 queue with Poisson arrival rate $a$ after forwarding and service time $S$ drawn from the request mix. The Pollaczek--Khinchine formula~\cite{harchol2013performance} gives $\mathbb{E}[T_q] = a\,\mathbb{E}[S^2]/\bigl(2(1-\rho)\bigr)$ at utilization $\rho = a\,\mathbb{E}[S]$. A KV cache match of $m$ tokens saves $\beta m$ of prefill time. Equating these costs gives the crossover utilization $\rho^*$ above which an idle worker that recomputes can outperform a loaded holder, motivating the score's capacity price $\lambda$.

Figure~\ref{fig:mg1} sweeps offered load in simulation on Mixed. Random forwarding follows the model despite a service-time $C_v^2$ of $5.4$, supporting the approximation. The predicted crossover is $\rho^*=0.13$ to $0.37$ for matches of 1K to 4K tokens, suggesting that load matters well below saturation. Let $\lambda_0$ denote the no-reuse capacity, the arrival rate at which offered work equals the pool's service rate with every prefix recomputed in full. Forwarding to the least loaded worker removes nearly all queueing below this point, with no measurable waits below $0.5\,\lambda_0$, but its knee remains at $\lambda_0$. The full score reduces mean residual prefill by 42\% and shifts the knee to about $1.65\,\lambda_0$, consistent with $1/(1-0.42)\approx 1.7$. Load awareness reduces queueing, while KV cache reuse reduces work and increases capacity.

\noindent\textbf{Index write cost.}
Each KV cache state event edits at most $L$ entries, one per indexed length covered by the prefix. Each edit takes amortized $O(1)$ work at its maintenance unit and produces a $(\mathrm{bucket}, \mathrm{slot}, \mathrm{value})$ triple with a 32-bit or 64-bit value, or a short cuckoo-move path for a new key. Propagation therefore sends at most $L$ updates per event to each edge, independent of $E$.

\noindent\textbf{Migration cost.}
A KV cache replica can be produced by transferring its tensors over the network or shipping the prompt and recomputing them at the destination. The comparison weighs network transfer time against the GPU time needed to recreate the same prefix. With $s$ KV bytes per token and prefill time $\beta$ ms/token, the model favors tensor transfer when bandwidth in bytes/ms exceeds $s/\beta$. Regional networks provide tens to hundreds of Gbps and typically exceed this threshold, while inter-region links over the public Internet fall below it, so \sys transfers KV cache replicas within regions and recomputes shipped prefixes across regions. Prompts contain only a few bytes per token, making their transfer cost small. Registration uses an \texttt{insert}; migration targets idle workers to limit interference with serving requests.

\noindent\textbf{Queue-view propagation traffic.}
A region's queue vector holds one $b$-bit estimate per worker, giving an average vector size of $(W/R)\,b$ bits. Under broadcast alone, each region sends its vector to the other $R-1$ regions every $\Delta$, averaging $(R-1)(W/R)\,b$ bits per region per period. Across the pool, this costs $(R-1)\,W b/\Delta$ bits/s, including unequal regional worker counts. Piggybacking carries vectors on existing WAN responses, and suppressing broadcasts to regions that received a piggybacked vector in the current period can reduce traffic below this bound.

%% file: 03_implementation.tex
\section{Implementation}
\label{sec:impl}

As shown in \autoref{fig:implementation}, we implement \sys in two forms: a real-machine prototype for end-to-end validation and an event-driven simulator for large-scale evaluation. The real-machine prototype couples instrumented SGLang workers with the KV cache discovery plane, while the simulator reuses the discovery-plane logic and models workers and network execution as events.

\noindent\textbf{Real-machine prototype.} The prototype consists of SGLang workers and the KV cache discovery plane.
SGLang workers are built on SGLang v0.5.10.
Each worker serves one model replica deployed on one or more GPUs, provides request scheduling and model execution, manages the two-tier KV cache through admission, demotion, promotion, and eviction, and exposes a control module for load reporting, KV cache state-event reporting, and replica migration coordination.
Cross-region requests are forwarded over the real WAN to the selected worker.

The discovery plane builds on the reference C++ Ludo implementation~\cite{ludo}  and comprises multi-level lookup units at edge nodes, sharded maintenance units, and the joint-score forwarding logic. Maintenance units process worker-reported KV cache events and propagate the resulting state updates to the lookup units. Edge nodes then rank workers with the joint score (Section~\ref{sec:design:ops}) under their local view.

\noindent\textbf{Simulator.}
The simulator reuses the same lookup, maintenance, and cache-state logic, while replacing physical workers and the real network with their event-driven counterparts. Its request loop advances a virtual clock, replays requests in arrival order, and, for each request, models forwarding, queueing, service, and subsequent KV cache state updates. The simulated worker pool captures queueing delay, prefill and decode time, inter-region network latency, and the two-tier VRAM and DRAM KV cache, with prefill, decode, and WAN costs calibrated to hardware benchmarks and network measurements. This calibration and the shared discovery-plane logic connect the simulator to the prototype, but we have not quantified simulation error against a testbed run under identical settings.

%% file: 04_eval.tex
\begin{figure*}[t]
  \centering
  \includegraphics[width=\linewidth]{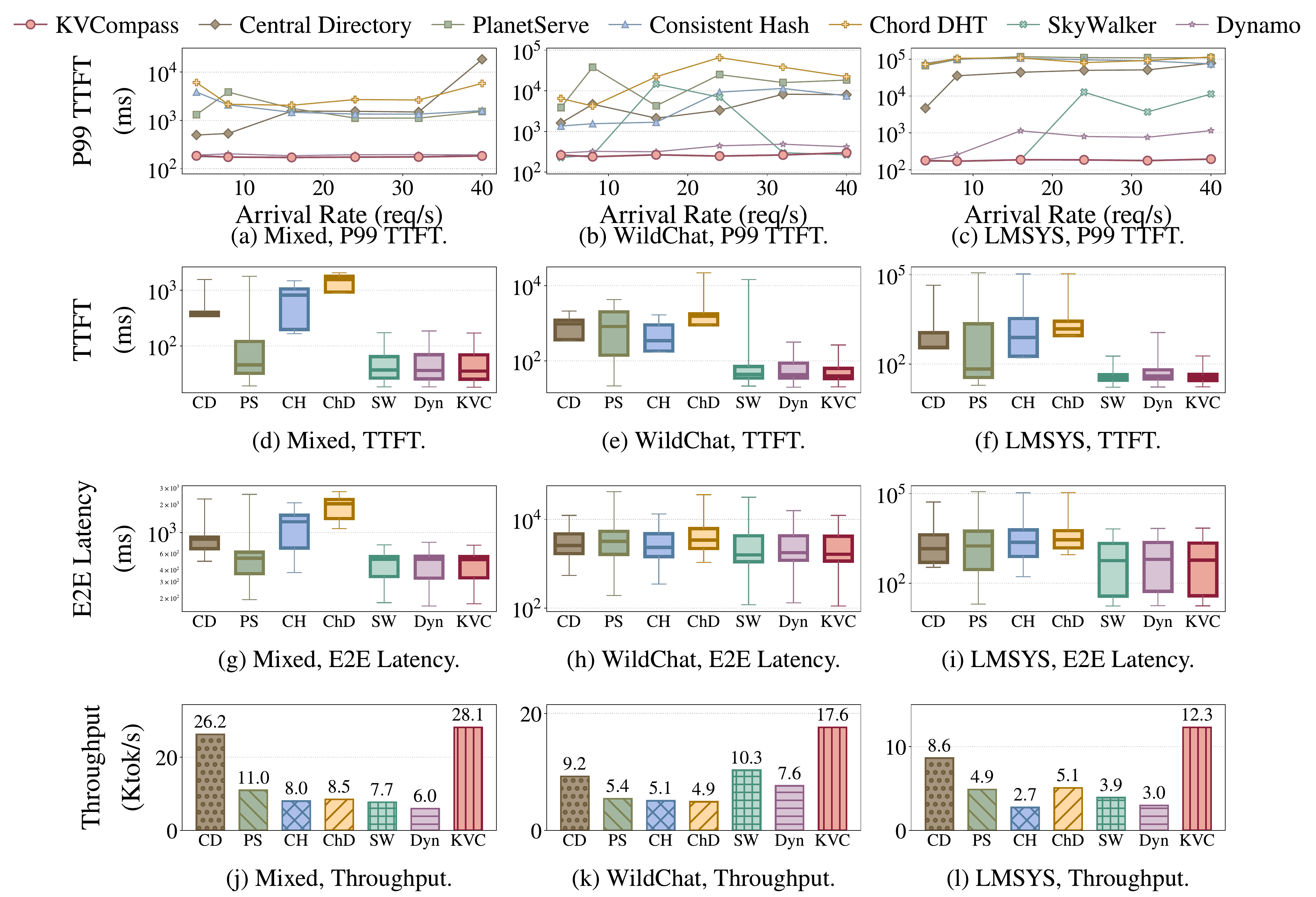}
  \caption{P99 TTFT vs.\ arrival rate, TTFT and end-to-end latency distributions, and throughput on Mixed, WildChat, and LMSYS, measured on the 14 workers across three-continent real-machine testbed (8/2/4 workers in US/EU/Asia) over the public Internet.}
  \label{fig:latency-real}
  \Description{Twelve panels compare seven schemes on the three-continent testbed. Rows show P99 TTFT versus arrival rate, TTFT distributions, end-to-end latency distributions, and throughput for Mixed, WildChat, and LMSYS.}
\end{figure*}

\begin{figure*}[t]
  \centering
  \includegraphics[width=\linewidth]{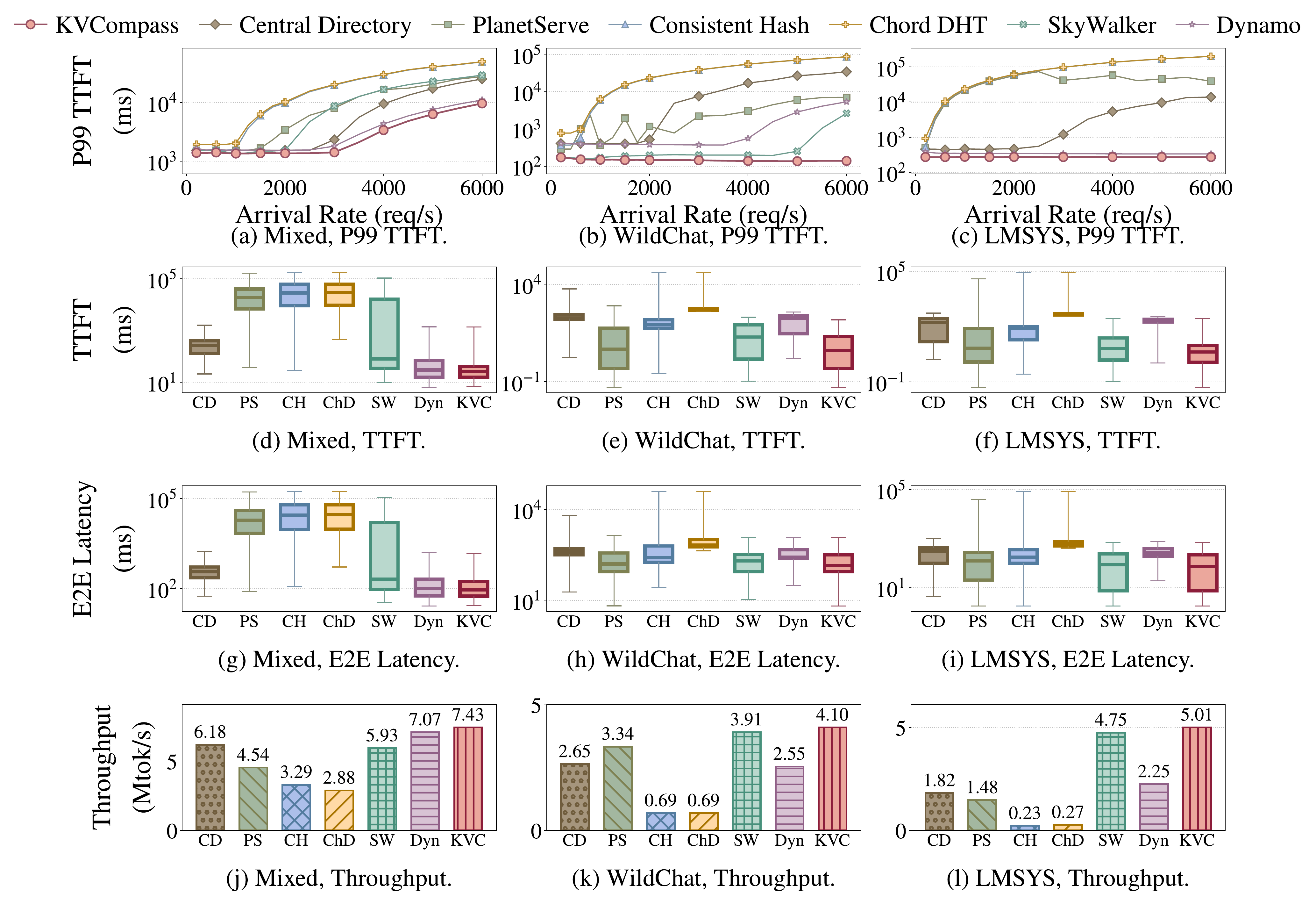}
  \caption{P99 TTFT vs.\ arrival rate, TTFT and end-to-end latency distributions, and throughput on Mixed, WildChat, and LMSYS, simulated on 64 workers across 8 regions.}
  \label{fig:latency-sim}
  \Description{Twelve panels compare seven schemes in the 64-worker, eight-region simulation. Rows show P99 TTFT versus arrival rate, TTFT distributions, end-to-end latency distributions, and throughput for three workloads.}
\end{figure*}

\section{Evaluation}
\label{sec:eval}
In this section, we evaluate TTFT latency, end-to-end latency and end-to-end service throughput both on the real-machine testbed (Section~\ref{sec:eval:real}) and large-scale simulation (Section~\ref{sec:eval:sim}).
We also evaluate load balancing, reported as each worker's load relative to the mean load across all workers. We then quantify the contribution of each mechanism by ablation study (Section~\ref{sec:eval:ablation}), measure the cost of every index operation (Section~\ref{sec:eval:micro}), and measure index memory
usage (Section~\ref{sec:eval:memory}). Appendix~\ref{app:scale} reports larger-scale results with 128 workers across 16 regions at arrival rates up to 18000\,req/s. Appendix~\ref{app:na} reports additional testbed results across four North-American regions.

\subsection{Experimental Methodology}
\noindent\textbf{Workload.}  We evaluate on three different workloads to reflect different deployment scenarios. \emph{Mixed} consists of three datasets: ToolBench~\cite{toolbench} for agent tool-calling traces (50\%), LooGLE~\cite{loogle} for long-document QA (5\%) and APPS~\cite{APPS} for competitive-programming problems (45\%). \emph{WildChat}~\cite{zhao2024wildchat} consists of real user conversations collected from ChatGPT. \emph{LMSYS}~\cite{zheng2024lmsyschatm} consists of real-world conversations collected from the Chatbot Arena platform.

\noindent\textbf{Eviction policy.} The discovery plane does not control each worker's local KV cache eviction policy. The results of \autoref{fig:evict} show that the best eviction policy differs across different workloads. To isolate discovery and forwarding gains from eviction effects, we use LFU for Mixed and LRU for WildChat and LMSYS, the best policy for each workload.

\noindent\textbf{Baselines.} We compare \sys with six baselines: (1) Consistent Hashing, using six independent content-hash rings; (2) Chord~\cite{chord}, using the six-level hash-based layout with one DHT key per chunk-size level; (3) PlanetServe~\cite{planetserve} using hash radix tree; (4) Central Directory, a centralized coordinator with global KV cache state; (5)~SkyWalker~\cite{SkyWalker}, which deploys a load balancer on each region and forwards a request to a replica only when the replica's pending queue is empty using ``selective pushing'' mechanism,  and (6)~Dynamo~\cite{nvidia-dynamo}, which deploys a KV cache-aware dynamo router implemented by Nvidia on each region and periodically broadcasts to get the global view.

\noindent\textbf{Configuration.} The testbed and the simulator share the same configurations. The queue-estimate broadcast period is $\Delta = 100$\,ms with EWMA weight $w = 0.5$, and the capacity price is capped at $\lambda_{\max} = 8$. Migration triggers at $H = 3$ misses of a prefix family within a $T_h = 1$\,s sliding window.

\subsection{Testbed Validation}
\label{sec:eval:real}

\noindent\textbf{Testbed Setup.}
We use up to 14 workers on three Vast.ai\footnote{\url{https://vast.ai}} multi-GPU servers across three regions, with 8 RTX~5090 GPUs in the US, 4 in Asia, and 2 in Europe. Each GPU hosts one SGLang~\cite{sglang} worker serving Llama-3.1-8B-Instruct (BF16). Within each region, the edge dispatches requests over kernel loopback TCP. Across regions, all traffic traverses the public Internet, with measured one-way latencies of 160\,ms between US and Europe, 126\,ms between US and Asia, and 149\,ms between Europe and Asia. \sys derives indexed lengths at startup from the calibrated RTX~5090 prefill speed of 0.085\,ms/token.

\noindent\textbf{P99 TTFT under increasing load.} Figure~\ref{fig:latency-real} (a)(b)(c) show P99 TTFT at arrival rates of 4--40\,req/s. At 16\,req/s, \sys is \textbf{8.6--12$\times$} lower than the hash-placement schemes and the Central Directory on Mixed, \textbf{6.3--83$\times$} lower on WildChat, and \textbf{236--626$\times$} lower on LMSYS. Hash-placement schemes route by prefix hash, concentrating popular-prefix requests on a few workers and overloading them. The Central Directory pays a WAN round trip to a single coordinator per request, increasing TTFT. The other two baselines match \sys at low load but degrade as the rate increases. SkyWalker's P99 TTFT spikes to \textbf{55$\times$} above \sys on WildChat at 16\,req/s and \textbf{69$\times$} on LMSYS at 24\,req/s. SkyWalker considers only queue occupancy, sending requests away from KV cache holders to pay full prefill time elsewhere. Dynamo is $4$--$6\times$ above \sys on LMSYS from 16\,req/s onward because its broadcast global view becomes stale between refreshes.

\noindent\textbf{Latency distributions.} Figure~\ref{fig:latency-real} (d)(e)(f) show TTFT latency box plots at 16\,req/s arrival rate. On WildChat, SkyWalker's P99 TTFT is \textbf{55$\times$} higher. Dynamo stays much closer, \textbf{1.2$\times$} higher. For end-to-end latency at 16\,req/s rate, Figure~\ref{fig:latency-real}(g)(h)(i) show that \sys's P99 is \textbf{3.5$\times$} lower than PlanetServe, \textbf{3.7$\times$} lower than Chord, and \textbf{3.1$\times$} lower than the Central Directory on Mixed. 

\noindent\textbf{Service throughput.} We show the service throughput in Figure~\ref{fig:latency-real} (j)(k)(l). \sys keeps the highest throughput on all three workloads. On Mixed, the Central Directory comes within 7\% of \sys. The other baselines deliver \textbf{2.6--4.7$\times$} lower throughput. On WildChat, \sys delivers \textbf{1.7$\times$} the throughput of SkyWalker, and \textbf{1.9--3.6$\times$} that of the rest. On LMSYS, \sys delivers \textbf{1.4$\times$} the throughput of the Central Directory and \textbf{2.4--4.5$\times$} that of the rest. Every recomputed prefix consumes GPU time that could have served new requests, and concentrating load onto a few workers wastes the computation of the idle rest. \sys jointly considers KV cache reuse and load across all workers, achieving the highest throughput.

\begin{figure}[t]
    \begin{minipage}[c]{0.48\linewidth}
        \centering
        \includegraphics[width=\linewidth]{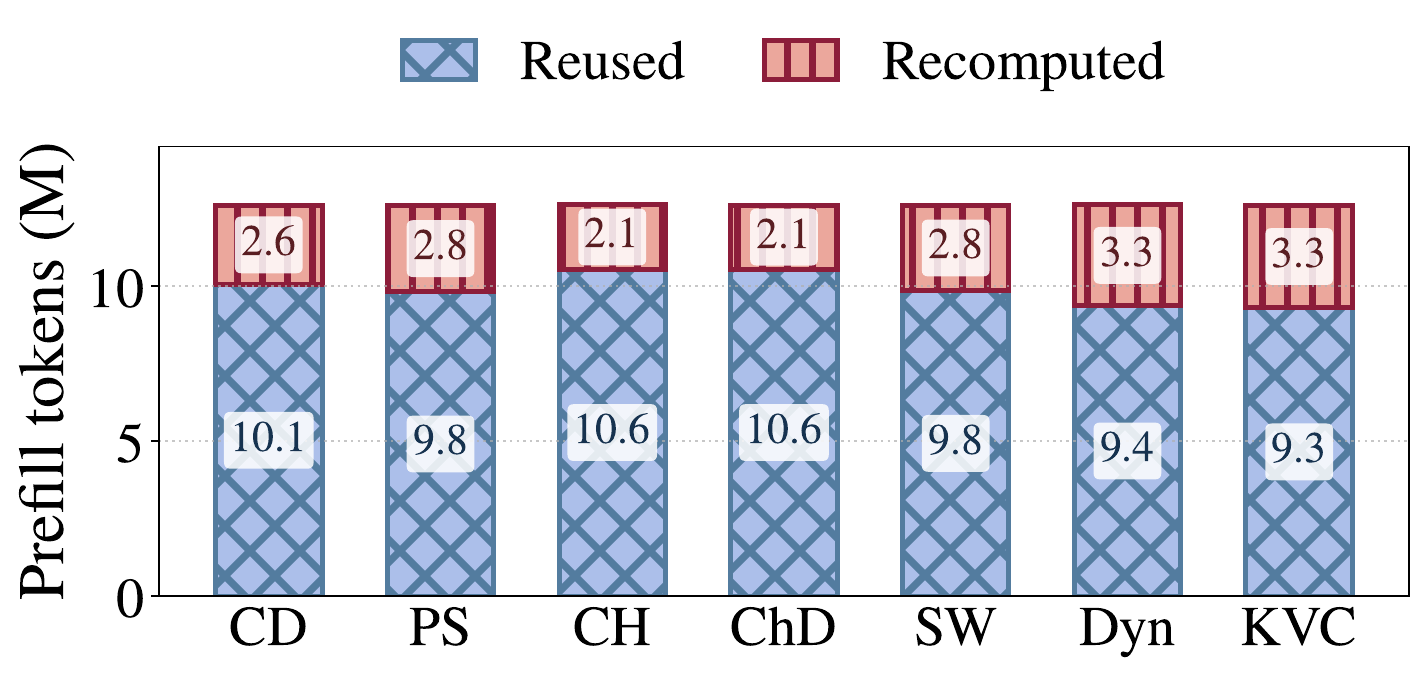}
        \caption{Reused versus recomputed prefill tokens by scheme on the real-machine testbed (Mixed).}
        \label{fig:reuse}
        \Description{Stacked bars show reused and recomputed prefill tokens for seven schemes on Mixed. Hash-placement schemes reuse more tokens than KVCompass.}
    \end{minipage}\hfill
    \begin{minipage}[c]{0.48\linewidth}
        \centering
        \includegraphics[width=\linewidth]{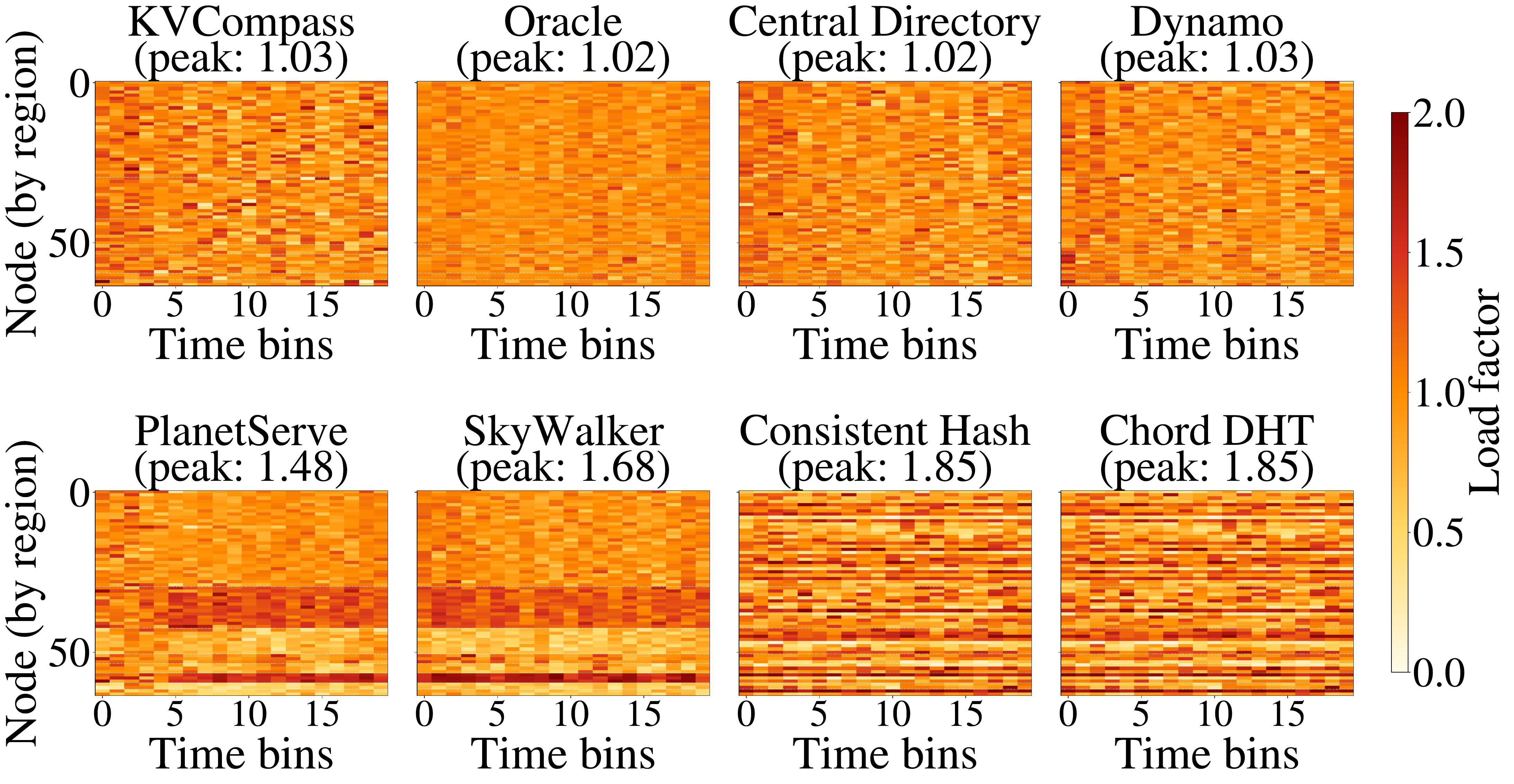}
        \caption{Per-worker load factor heatmap across schemes. Peak: the hottest
        worker's time-averaged load factor, relative to each scheme's balanced mean.}
        \label{fig:heatmap}
        \Description{Heatmaps show per-worker load factors over time for seven schemes and the Oracle. KVCompass has a peak factor of 1.03, close to the Oracle peak of 1.02.}
    \end{minipage}
\end{figure}

\noindent\textbf{Reused versus recomputed tokens.} \autoref{fig:reuse} shows each scheme's prefill tokens split into those reused from a KV cache hit and those recomputed from scratch. \sys does not reuse the most. The hash-placement schemes reuse the largest share, 10.6\,M tokens against \sys's 9.3\,M on Mixed, because they always route a prefix to the same worker, and \sys recomputes somewhat more. \sys nonetheless delivers the highest throughput of all schemes (\autoref{fig:latency-real}), so its gains come mainly from load-aware forwarding rather than maximizing cache reuse.

\subsection{Large-Scale Simulation}
\label{sec:eval:sim}
We scale the same Mixed workload to 50 independent tenants each replaying the workload with a unique private system prompt, producing $\approx$375\,K distinct prefix families across 64 workers in 8 geographic regions. Per-region capacity follows a real LLM-deployment
census~\cite{llm-deployment-wild}, so the worker distribution matches a measured deployment, giving 30 workers in the United
States, 13 in China, 8 in Germany, and 2--3 in each of Japan,
Singapore, the United Kingdom, South Korea, and Australia. The 30-worker United States region uses 64-bit discovery index values, while the other regions use 32-bit values.
Each worker is modeled as an A100-80GB GPU running Llama-3.1-8B-Instruct~\cite{llama}(BF16), with 32\,GB VRAM and 128\,GB host DRAM reloadable over PCIe 4.0 at $\approx$25\,GB/s. Inter-region WAN latency follows a realistic matrix (12--150\,ms RTT) derived from CloudPing.co AWS measurements, the wide-area latencies a real inter-region deployment would see. Prefill and decode costs are calibrated against published A100 BF16 benchmarks on vLLM~\cite{vllm} and SGLang~\cite{sglang}, so per-token compute times match real hardware. Requests are replayed at 200--6000\,req/s. We compare to the same six baselines as the real-machine testbed.

\noindent\textbf{Latency and throughput.} Figure~\ref{fig:latency-sim} plots P99 TTFT versus arrival rate and latency distribution at 2750\,req/s. At 3000\,req/s on Mixed, \sys achieves P99 TTFT \textbf{1.6$\times$} lower than the Central Directory, \textbf{5.7$\times$} lower than PlanetServe, \textbf{6.2$\times$} lower than SkyWalker, \textbf{14$\times$} lower than the DHT schemes, and \textbf{1.3$\times$} lower than Dynamo. Under sustained overload at 6000\,req/s the advantage holds at \textbf{1.1--5.1$\times$} across all baselines. The box panels show that the advantage scales to 64 workers: across the three workloads at 2750\,req/s, the DHT schemes land \textbf{125--312$\times$} above \sys on P99 TTFT and \textbf{33--118$\times$} on P99 end-to-end latency, PlanetServe lands \textbf{5.2--145$\times$} and \textbf{1.2--111$\times$} above, and the Central Directory lands \textbf{1.2--38$\times$} and \textbf{1.2--5.4$\times$} above. Among the load-aware baselines, SkyWalker sits \textbf{78$\times$} above \sys on P99 TTFT on Mixed and Dynamo \textbf{2.5$\times$} above on WildChat. The throughput panels show that \sys delivers the highest throughput on all three workloads: about \textbf{5\%} above
Dynamo and SkyWalker, and \textbf{2.3--22$\times$} above the hash-placement
schemes. \sys keeps both its latency and its throughput advantages in large-scale simulation.

\noindent\textbf{Load Balancing.}
A worker's load factor measures how far it sits from an even split of the load. Over a time window, we sum the GPU compute time of the requests sent to each worker and divide it by the average over all workers. \autoref{fig:heatmap} maps each worker's load factor across the simulated 64 workers over a 30-second window at the 2750\,req/s arrival rate on the Mixed workload. Each scheme's peak is its hottest node's time-averaged load factor relative to its balanced mean.

The Oracle forwards every request to the least-loaded worker in the whole worker pool, so its peak of \textbf{1.02} is the best lower bound any scheme can achieve. \sys peaks at \textbf{1.03}, within \textbf{2\%} of the Oracle. Our discovery plane considers the current estimated queueing time to every worker globally. 
The Central Directory and Dynamo balance comparably, peaking at 1.02 and 1.03, since the Central Directory holds exact global state and the Dynamo routers on each region are load-aware.
SkyWalker peaks at 1.68, since its locality-first policy keeps every
request inside its local region as long as any local replica is free.
The hash-placement schemes peak at 1.85 and PlanetServe at 1.48, concentrating the requests for popular prefixes on a few hash-designated workers, the same failure seen in Section~\ref{sec:eval:real}. Read together with the reuse breakdown of \autoref{fig:reuse}, this suggests that in these experiments a better load balance that avoids these hot spots contributes more to performance than maximizing reused KV cache does.

\subsection{Ablation Study}
\label{sec:eval:ablation}

\begin{wrapfigure}{r}{0.44\textwidth}
    \centering
    \includegraphics[width=0.42\textwidth]{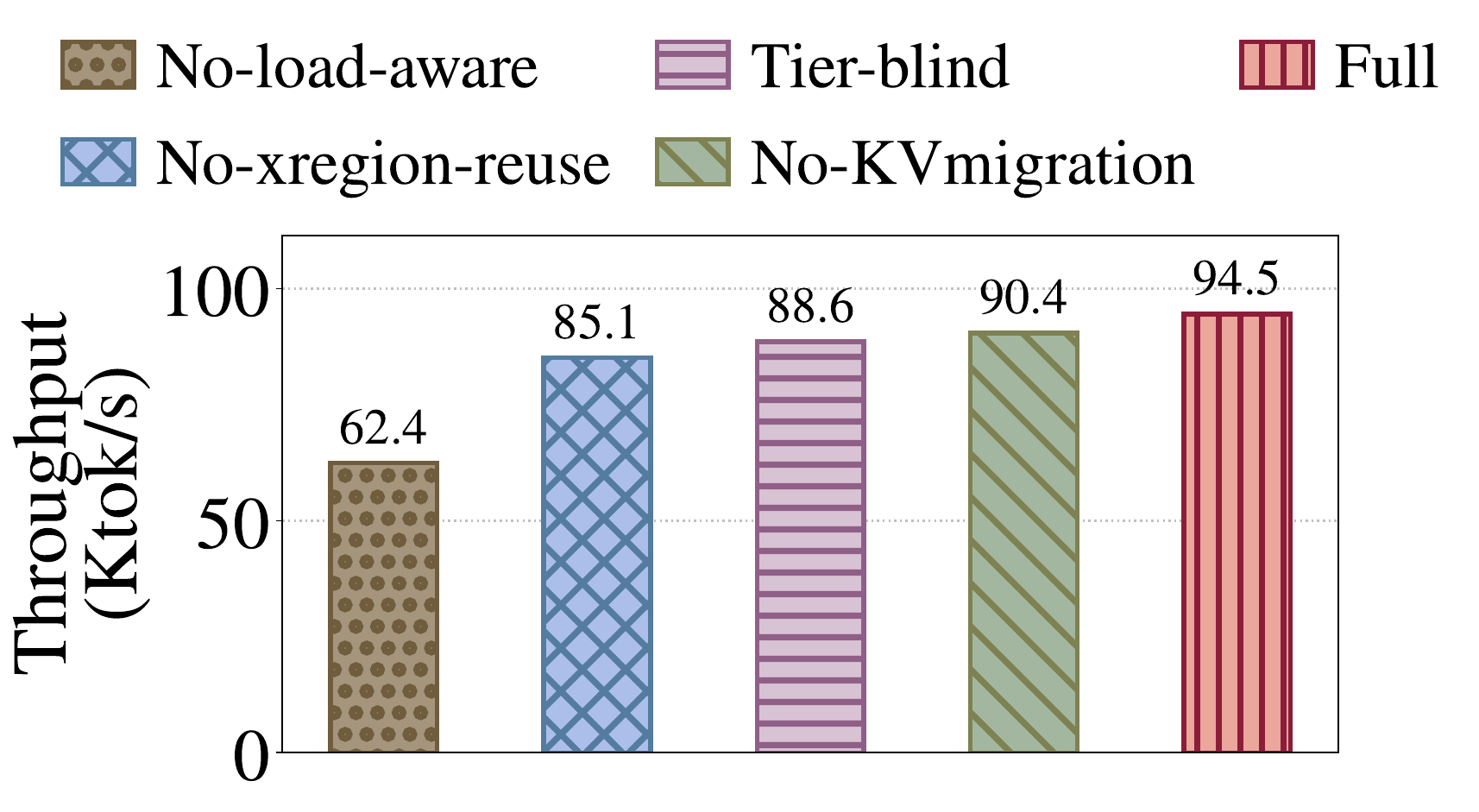}
    \caption{Ablation study. Service throughput with one mechanism removed at a time on the real-machine testbed (Mixed workload).}
    \label{fig:ablation}
    \Description{Bar chart of service throughput with each KVCompass mechanism disabled. Removing load awareness has the largest effect, followed by cross-region reuse, tier information, and KV cache migration.}
\end{wrapfigure}

In order to evaluate each mechanism of \sys and quantify their contribution, we disable one mechanism at a time and measure closed-loop service throughput under the same real-machine testbed setting, on the Mixed workload.

\autoref{fig:ablation} shows the results. No-loadaware removes estimated queueing time from the joint score, leaving only KV cache affinity and network latency. Throughput drops by 34\% as requests concentrate on workers holding hot-prefix KV caches. No-xregion-reuse disables cross-region lookups and forwarding. Throughput drops by 10\% because requests recompute prefixes locally even when remote KV cache reuse would save more time than the added network latency. Tier-blind removes KV cache tier information from the discovery index, pricing DRAM hits like VRAM hits. Throughput drops by 6\% because ignoring DRAM reload time can favor slower workers. No-KVmigration disables proactive migration of hot KV caches. Throughput drops by 4\% as requests either recompute missing prefixes or pay extra network latency to reach a holder.

\subsection{Operations Microbenchmarks}
\label{sec:eval:micro}
We benchmark discovery index operations on a dual 96-core CPU host of the real-machine testbed. Table~\ref{tab:micro} reports P50 and P99 latency and single-thread throughput, with write operations measured at 85--95\% table occupancy.

\noindent\textbf{Lookup and Forwarding.}
A lookup in a view of $10^7$ entries takes 310\,ns at the P50 and 590\,ns at the P99. For a 4096-token prompt, forwarding queries a five-level Ludo table and scores all 14 workers, taking 20.6\,$\mu$s at the P50 and 21.5\,$\mu$s at the P99, with 48\,K decisions per second on one thread. These costs add little overhead to request serving.

\noindent\textbf{Insert.}
P50 insert latency rises from 1.2\,$\mu$s at 85\% occupancy to 1.4\,$\mu$s at 95\%, while P99 rises from 4.4 to 6.1\,$\mu$s. Fuller tables increase the chance of hitting occupied slots and displacing entries.

\noindent\textbf{Update.}
An update rewrites one 2-bit field in the prototype's 32-bit value. It is the cheapest write operation, taking 170\,ns at the P50 across the tested occupancies. P99 latency is 1.0--1.8\,$\mu$s, with throughput of 4.8--5.8\,M updates per second.

\noindent\textbf{Remove.}
P50 remove latency is 0.8--0.9\,$\mu$s, close to insertion because clearing a slot follows the same path as writing it. P99 latency is 4.8--5.9\,$\mu$s, with throughput of 0.8--0.9\,M removals per second. Table~\ref{tab:micro} gives the remaining latency and throughput results.

%% file: 05_ablation.tex

\begin{table}[t]
    \begin{minipage}[t]{0.48\linewidth}
        \centering
        \small
        \caption{Cost of each discovery index operation. Write-path ranges span 85--95\% table occupancy.}
        \label{tab:micro}
  \begin{tabular}{@{}lrrr@{}}
            \toprule
            Operation & P50 Time & P99 Time & Rate \\
            \midrule
            Lookup            & 310\,ns & 590\,ns & 5.4\,M/s \\
            Forwarding        & 20.6\,$\mu$s & 21.5\,$\mu$s & 48\,K/s \\
            Insert            & 1.2--1.4\,$\mu$s & 4.4--6.1\,$\mu$s & 0.6--0.7\,M/s \\
            Update            & 170\,ns & 1.0--1.8\,$\mu$s & 4.8--5.8\,M/s \\
            Remove            & 0.8--0.9\,$\mu$s & 4.8--5.9\,$\mu$s & 0.8--0.9\,M/s \\
            \bottomrule
        \end{tabular}
    \end{minipage}\hfill
    \begin{minipage}[t]{0.48\linewidth}
        \centering
        \small
        \caption{Discovery index memory at $10^9$ entries.}
        \label{tab:memory}
  \begin{tabular}{@{}lrr@{}}
            \toprule
            Data Structure & Bits/entry & Total \\
            \midrule
            \sys                   & 48  & 6.0\,GB \\
            Hash map directory     & 498 & 62\,GB \\
            Token radix tree       & 12{,}900 & 1.6\,TB \\
            PlanetServe radix tree & 73 & 9.2\,GB \\
            \bottomrule
        \end{tabular}
    \end{minipage}
\end{table}
\subsection{Discovery Index Memory Footprint}
\label{sec:eval:memory}
Table~\ref{tab:memory} compares memory footprints at $10^9$ entries. The benchmark uses 14 workers and 32-bit values for \sys, with real Mixed token sequences at smaller scales. At $10^9$ entries, we directly allocate and write to the lookup view's bucket and classifier arrays, measuring \textbf{48 bits per entry} and \textbf{6.0\,GB} without inserting all prefixes. The difference from the theoretical 37 bits per entry includes allocation overhead. The hash map and token radix tree estimates are extrapolated from smaller runs, giving 498 bits per entry and 1.6\,TB, respectively. PlanetServe's hash radix tree uses 9.2\,GB in a full-scale run. These \sys results apply to the measured 32-bit layout. Tables with 64-bit values use more memory per entry, and total memory depends on the entry count and allocation overhead of all tables.
\subsection{Robustness to WAN and Control-Plane Variation}
\label{sec:eval:robust}
We evaluate WAN latency jitter, discovery index update loss, and update delay in the large-scale simulation on Mixed at 3000\,req/s. Actual WAN latency varies around the static latency matrix used by the joint score. Table~\ref{tab:robust} shows that $\pm$100\% jitter preserves the KV cache hit rate and changes P99 TTFT by less than 2\%. With 30\% update loss, the hit rate drops by 2.5 percentage points and P99 TTFT rises by 4.2\%. A 300\,ms update delay changes the hit rate by 0.3 percentage points and P99 TTFT by 3.4\%. Inaccurate latency estimates and stale entries may reduce reuse or increase latency, but the selected worker uses only the KV cache it holds and recomputes the rest. The prototype does not automatically retransmit failed update batches. Later worker reports or reloading the maintenance units' lookup view may correct stale entries, with no guaranteed recovery time.

\subsection{False Hits in the Keyless Discovery Index}
\label{sec:eval:falsehit}
The edge lookup view stores no keys, so a query for a never-inserted prefix can return a false locator from an occupied Ludo slot. Table~\ref{tab:falsehit} reports two rates. The structural false-positive rate measures how often a never-inserted key returns a false locator, reaching 45.3\% at 95\% table load. The end-to-end false-lookup rate measures how often the worker selected for a real request lacks the claimed KV cache match, giving 0.10\% on Mixed, 1.27\% on WildChat, and 0.34\% on LMSYS. These rates use different query populations and measure different stages. Real requests can query prefixes already in the discovery index, and a false locator affects forwarding only if its worker wins the joint score. Cache belief discounts the claimed matches of workers that repeatedly reuse fewer tokens than claimed. A false lookup may reduce KV cache reuse, but the selected worker recomputes any missing prefix, preserving answer correctness.

\begin{table}[t]
  \centering
  \small
  \caption{\sys sensitivity to WAN jitter and update loss/delay (Mixed, 3000\,req/s).}
  \label{tab:robust}
  \begin{tabular*}{\linewidth}{@{\extracolsep{\fill}}lcccccccccc@{}}
    \toprule
     & Base & \multicolumn{3}{c}{Jitter ($\pm$\%)} & \multicolumn{3}{c}{Loss (\%)} & \multicolumn{3}{c}{Delay (ms)} \\
    \cmidrule(lr){3-5}\cmidrule(lr){6-8}\cmidrule(lr){9-11}
     & 0 & 25 & 50 & 100 & 10 & 20 & 30 & 100 & 200 & 300 \\
    \midrule
    P99 TTFT (ms) & 1370 & 1355 & 1345 & 1345 & 1328 & 1404 & 1427 & 1343 & 1324 & 1324 \\
    Hit rate (\%) & 43.8 & 43.8 & 43.8 & 43.8 & 42.7 & 42.3 & 41.3 & 42.9 & 43.4 & 44.1 \\
    \bottomrule
  \end{tabular*}
\end{table}

%% file: 06_discussion.tex
\section{Discussion}
\label{sec:discussion}
\noindent\textbf{Serving multiple models.} Our main evaluation serves one model across all workers, while Internet-scale deployments may serve several models. Even for the same prefix, different models, precisions, or parallel partitions produce incompatible KV tensors~\cite{promptcache,cachegen}. Including model identity and configuration in the discovery index key would keep these KV caches separate. A lookup would then return only workers holding KV caches compatible with the request's model and configuration. Entries for different models could coexist in one table without changing the discovery index structure, operations, or lookup views. The orchestration layer would still decide which workers load each model and when to replace its weights~\cite{prism,serverlessllm}.

\begin{table}[t]
  \centering
  \small
  \caption{False hits in the keyless discovery index for never-inserted keys and real requests.}
  \label{tab:falsehit}
  \begin{tabular*}{\linewidth}{@{\extracolsep{\fill}}lcccccccc@{}}
    \toprule
     & \multicolumn{5}{c}{Never-inserted key, by table load} & \multicolumn{3}{c}{End-to-end, by workload} \\
    \cmidrule(lr){2-6}\cmidrule(lr){7-9}
     & 20\% & 40\% & 60\% & 80\% & 95\% & Mixed & WildChat & LMSYS \\
    \midrule
    False-positive rate (\%) & 9.7 & 19.1 & 28.6 & 38.2 & 45.3 & 0.10 & 1.27 & 0.34 \\
    \bottomrule
  \end{tabular*}
\end{table}

\noindent\textbf{Prefill--decode disaggregation.} Our main evaluation runs prefill and decode on the same worker. Disaggregated systems separate the two phases because prefill is compute-bound and decode is bound by memory bandwidth~\cite{distserve,splitwise,mooncake}. In this setting, \sys routes prefill requests and tracks KV caches on prefill workers, since reuse saves prefill computation. Decode output becomes part of the next turn's prefix and enters the discovery index when that prefix is prefilled. The serving stack selects decode workers and transfers KV caches. Appendix~\ref{app:pd} evaluates this integration on a three-region testbed with SGLang and Mooncake, where \sys achieves lower median and P99 TTFT than all evaluated baselines. Decode scheduling, transfer scheduling, and pool sizing remain the responsibility of the disaggregated serving stack.

\noindent\textbf{Security and privacy.} KV cache reuse requires matching the entire indexed prefix, so a lookup does not return another user's unknown prompt content. However, timing differences can reveal whether a known or guessed prefix is cached~\cite{planetserve,cache-audit,earlybird}. Existing timing attacks extend guesses a few tokens at a time. In \sys, indexed prefixes span at least hundreds of tokens, which may make incremental guessing more difficult, although an attacker who knows a suspected prefix may still infer its presence. Including a user or user-group identity in the discovery index key would restrict KV cache sharing to that group. The provider could choose the sharing boundary, while timing side channels within each group may still need mitigations such as response padding~\cite{predictive-mitigation}.

%% file: 07_related.tex
\section{Related Work}
\label{sec:related}
\noindent\textbf{Geo-distributed LLM serving.}
KDN~\cite{llm-cdn} and Prefill-as-a-Service~\cite{prfaas} motivate KV cache sharing across datacenters. PlanetServe~\cite{planetserve} supports decentralized serving with a radix-tree index for regional prefix reuse, while SkyWalker~\cite{SkyWalker} balances load across regions. GORGO~\cite{gorgo} optimizes cross-region routing using KV cache locality, network latency, and queue state, an objective close to \sys's joint score. Its KV cache index estimates prefix availability without directly observing worker evictions. \sys instead tracks KV cache holders and storage tiers through KV cache events and replicates compact lookup views at every edge node. Our main contribution is how reusable KV caches are discovered across regions.

\noindent\textbf{Intra-cluster prefix caching.}
SGLang~\cite{sglang}, ChunkAttention~\cite{chunkattention}, and Hydragen~\cite{hydragen} reuse shared prefix KV cache, while CacheBlend~\cite{cacheblend} reuses KV cache for arbitrary chunks. Mooncake~\cite{mooncake}, MemServe~\cite{memserve}, and Infinite-LLM~\cite{infinitellm} pool KV cache across storage tiers. DualMap~\cite{dualmap}, Helix~\cite{helix}, Preble~\cite{preble}, LMCache~\cite{lmcache}, and LoongServe~\cite{loongserve} support KV cache reuse and scheduling within a cluster. \sys's joint score builds on prior KV cache-aware and load-aware routing, while its discovery plane finds KV caches across regions at prefix lengths chosen from network latency.

\noindent\textbf{Distributed naming and discovery.}
Chord~\cite{chord}, Kademlia~\cite{kademlia}, Pastry~\cite{pastry}, and IPFS~\cite{ipfs} map an item's identity to a hash-designated node that holds or routes to it. KV caches are created on the workers that serve requests, so their locations cannot be derived from prefix hashes. \sys decouples KV cache identity from placement by tracking holders wherever KV caches are created.

\noindent\textbf{Compact key-value lookup structures.}
Othello~\cite{othello} provides keyless $O(1)$ lookups, Ludo~\cite{ludo} adds dynamic updates, and Smash~\cite{smash} applies explicit-write lookup to distributed storage placement. Cuckoo filters~\cite{cuckoo-filter} address set membership. \sys tracks changing KV cache holders and storage tiers at latency-derived prefix lengths for Internet-scale LLM serving.

%% file: 08_conclusion.tex
\section{Conclusion}
\label{sec:conclusion}
This paper presents \sys, a KV cache discovery plane for Internet-scale prefix sharing in LLM serving. By decoupling KV cache identity from placement, \sys discovers reusable KV caches wherever workers create them. Hash-placement schemes cannot provide this decoupling because they determine KV cache placement from its identity. Compact local lookups and a joint score balance prefix reuse, worker load, and network distance. Real-machine experiments and large-scale simulations show that \sys achieves lower TTFT and end-to-end latency and higher throughput than all baselines, while maintaining good load balance.

%% file: 09_appendix.tex

\section{Simulation at Larger Scale}
\label{app:scale}
We extend the large-scale simulation of Section~\ref{sec:eval:sim} to 128 workers across 16 geographic regions, using the same workloads and baselines with 50 independent tenants per workload. Per-region capacity follows the same real LLM-deployment census~\cite{llm-deployment-wild}, giving 26 and 14 workers in two United States regions, 12 in China, 10 each in Germany and Ireland, 8 each in Japan and the United Kingdom, and 2 to 7 in each of Singapore, South Korea, India, Australia, Brazil, Canada, Italy, Bahrain, and South Africa. Each worker is modeled as an A100-80GB GPU running Llama-3.1-8B-Instruct~\cite{llama}~(BF16), with 32\,GB VRAM and 128\,GB host DRAM reloadable over PCIe 4.0 at $\approx$25\,GB/s. Across region WAN latency follows the same CloudPing.co-derived matrix extended to the added regions by great-circle distance, with one-way latencies spanning 9 to 144\,ms across the 16 regions. Requests are replayed at 400--18000\,req/s, scaled up with the doubled capacity.

\begin{figure}[htbp]
  \centering
  \includegraphics[width=\linewidth]{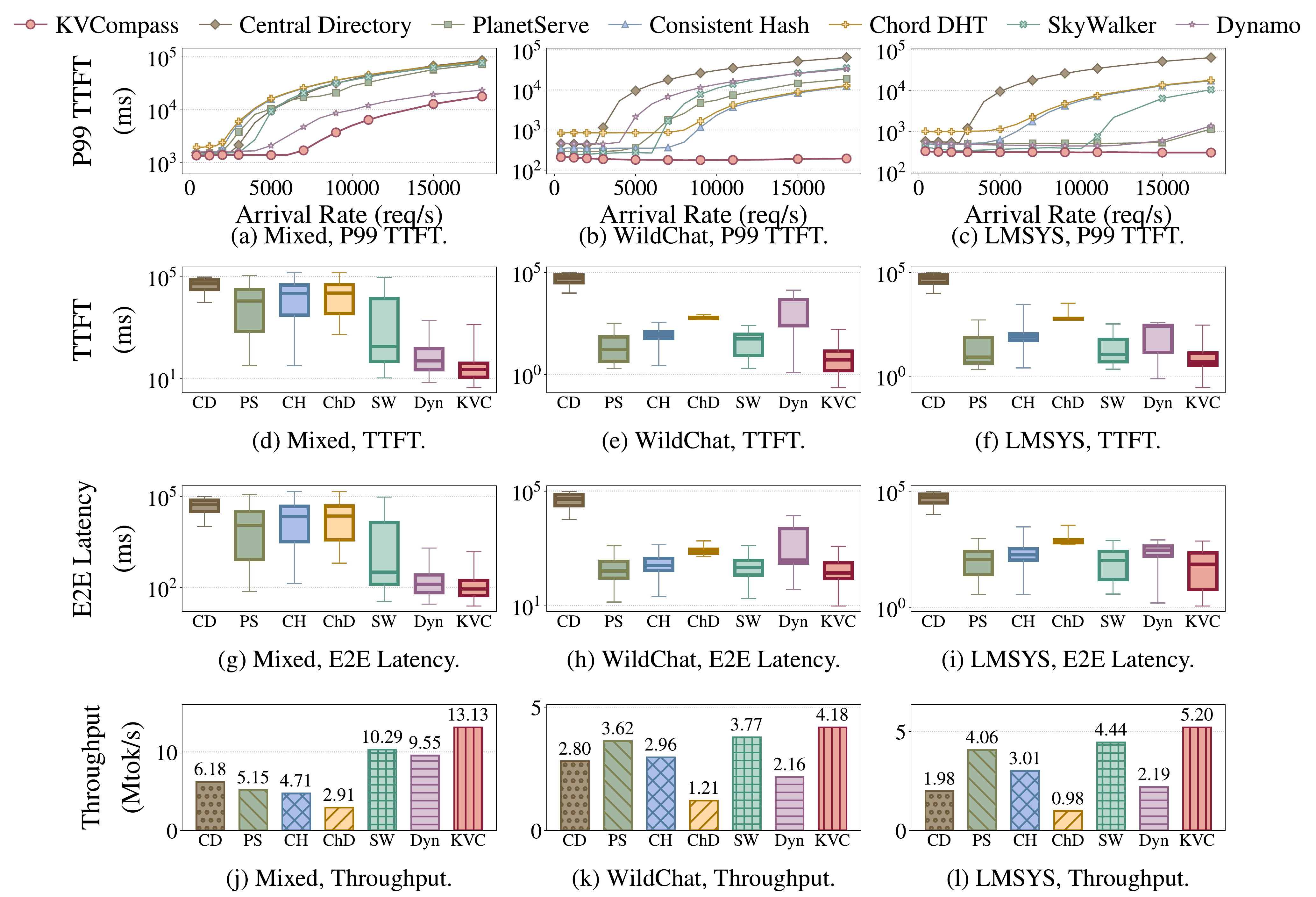}
  \caption{P99 TTFT vs.\ arrival rate, TTFT and end-to-end latency distributions, and throughput on Mixed, WildChat, and LMSYS, simulated on 128 workers across 16 regions.}
  \label{fig:sim-16region}
  \Description{Twelve panels compare seven schemes in the 128-worker, 16-region simulation, showing P99 TTFT versus arrival rate, TTFT and end-to-end latency distributions, and throughput across three workloads.}
\end{figure}

\noindent\textbf{Latency and throughput.} Figure~\ref{fig:sim-16region} plots P99 TTFT versus arrival rate and latency distribution at 5500\,req/s. Latency distributions use 500{,}000 requests per workload. At 12000\,req/s on Mixed, \sys achieves P99 TTFT \textbf{1.8--6.5$\times$} lower than every baseline; the gap widens to \textbf{27--218$\times$} on WildChat and \textbf{1.4--128$\times$} on LMSYS. \sys also delivers the highest throughput on all three workloads, 13.1, 4.2, and 5.2\,Mtok/s, \textbf{1.1--1.3$\times$} above the closest load-aware baseline and \textbf{1.4--5.3$\times$} above the hash-placement schemes. At twice the region count, worker count, and arrival rate of the main simulation, \sys still holds the best P99 TTFT, tail latency, and throughput on every workload, preserving its advantage at scale.

\section{Results over a Smaller Region Range}
\label{app:na}
\begin{figure*}[t]
  \centering
  \includegraphics[width=\linewidth]{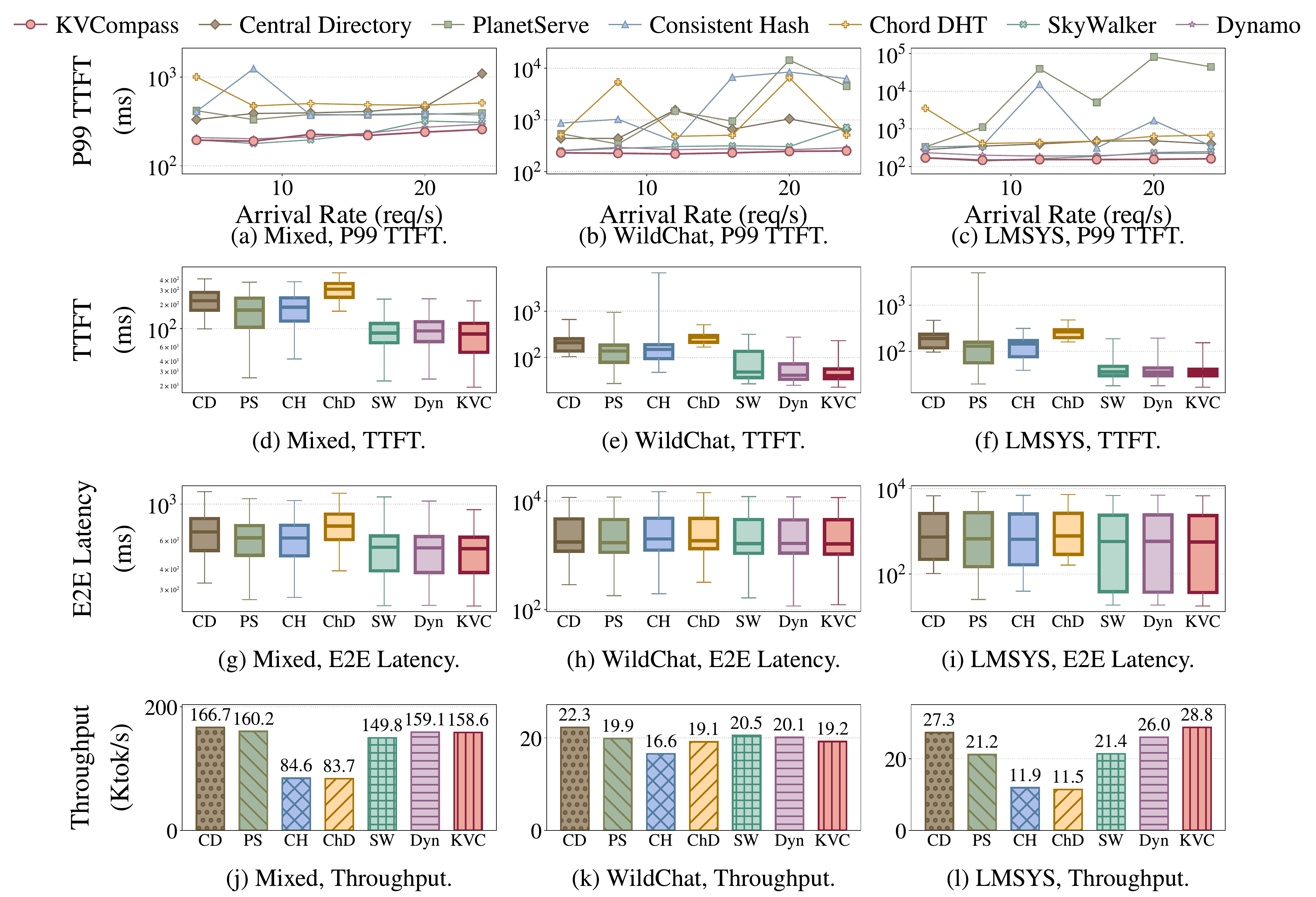}
  \caption{P99 TTFT vs.\ arrival rate, TTFT and end-to-end latency distributions, and throughput on Mixed, WildChat, and LMSYS, on an 11-worker, four-region North-American testbed over the public Internet.}
  \label{fig:latency-real-na}
  \Description{Twelve panels compare seven schemes on the 11-worker, four-region North-American testbed, showing P99 TTFT versus arrival rate, latency distributions, and throughput across three workloads.}
\end{figure*}

The testbed of Section~\ref{sec:eval:real} spans three continents. To check that the discovery plane holds up under a very different geography, we also ran the full evaluation on a second testbed over a much smaller region range. It uses 11 RTX~5090 workers across four North-American regions, with 2 GPUs in Pennsylvania, 4 in Ohio, 1 in New York, and 4 in Calgary, and one-way WAN latencies of 5.8 to 16.6\,ms among the three eastern regions and 24.9 to 37.4\,ms to Calgary, far shorter than the intercontinental links of Section~\ref{sec:eval:real}. Figure~\ref{fig:latency-real-na} reports the same three workloads and seven schemes. At 16\,req/s, \sys achieves P99 TTFT 1.7 to 2.2$\times$ below the hash-placement schemes and the Central Directory on Mixed and up to 29 to 33$\times$ on WildChat and LMSYS, and maintains competitive throughput. The advantage carries over to this much closer region range, which shows the design is robust and not tuned to one geography.

\section{Prefill-Decode Disaggregation}
\label{app:pd}
\sys assumes prefill and decode run on the same worker, but serving systems increasingly disaggregate the two phases onto separate pools~\cite{distserve,splitwise,mooncake}. \sys adapts naturally, because KV cache reuse saves prefill computation, so the discovery plane routes only the prefill request and indexes only the prefill workers, while the decode instance and the KV transfer are left to the disaggregated stack. We validate this on a three-region testbed of 7 prefill and 3 decode RTX~5090 instances, where the edge node plays SGLang's prefill-decode load balancer with Mooncake KV transfer~\cite{mooncake}. Every scheme selects the prefill worker, and the decode instance is drawn at random within the prefill's region identically for all schemes, so all separation comes from prefill routing. Figure~\ref{fig:pd} shows the TTFT CDF at 24\,req/s on the hot-skew Mixed workload against the stock prefill-decode stack with random dispatch, SGLang's built-in cache-aware router, round-robin, and consistent hashing. \sys has the lowest median TTFT, 1.25 to 1.69$\times$ below every baseline, and its P99 is 1.3 to 15.7$\times$ below, because random and hash dispatch let hot-prefix requests concentrate and their tails stretch past 9\,s.

\section{Cross-Region Reuse in Production Traces}
\label{app:traces}
\begin{figure}[t]
  \begin{minipage}[t]{0.48\linewidth}
    \centering
    \includegraphics[width=\linewidth]{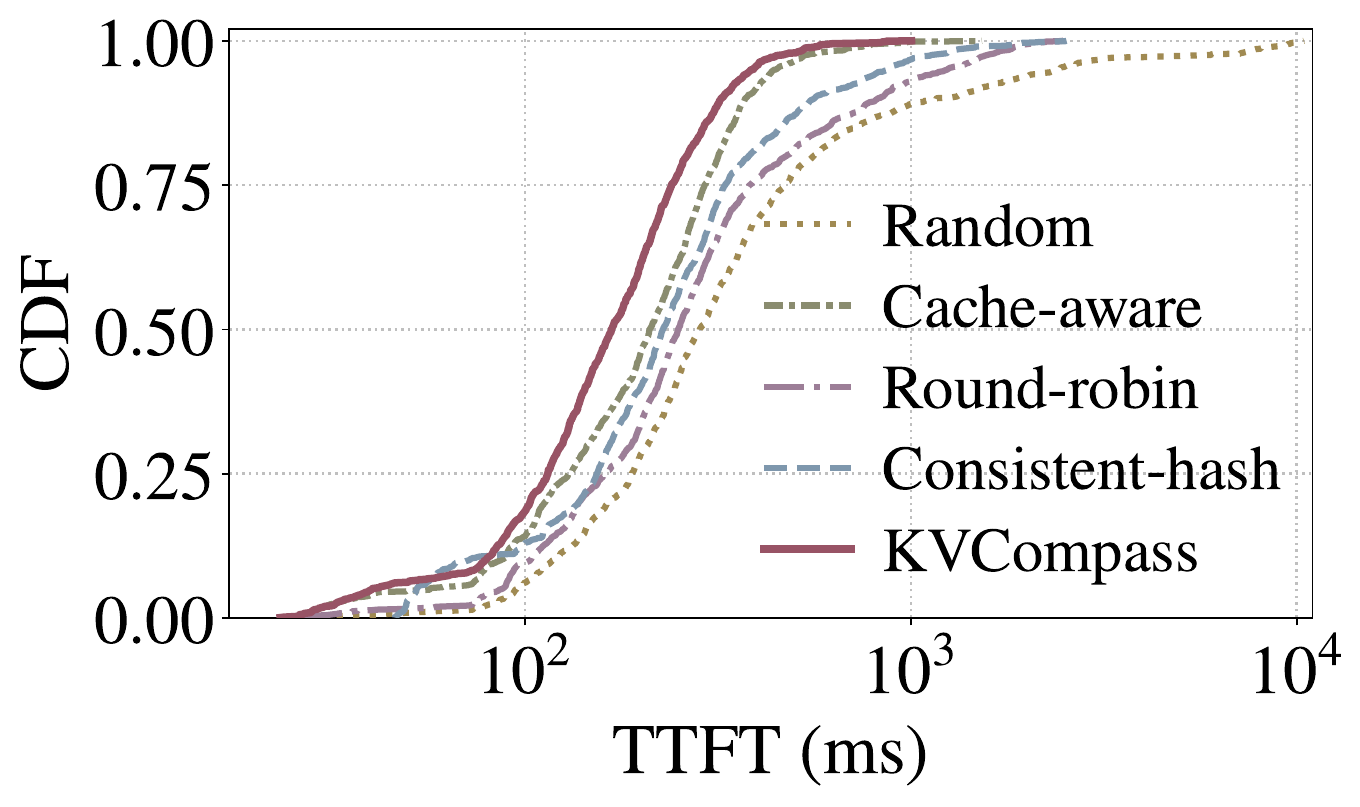}
    \caption{TTFT CDF under prefill-decode disaggregated serving (Mixed workload).}
    \label{fig:pd}
    \Description{CDF curves compare TTFT for KVCompass and four baselines under prefill-decode disaggregation on Mixed. KVCompass has lower median and P99 TTFT than the evaluated baselines.}
  \end{minipage}\hfill
  \begin{minipage}[t]{0.48\linewidth}
    \centering
    \includegraphics[width=\linewidth]{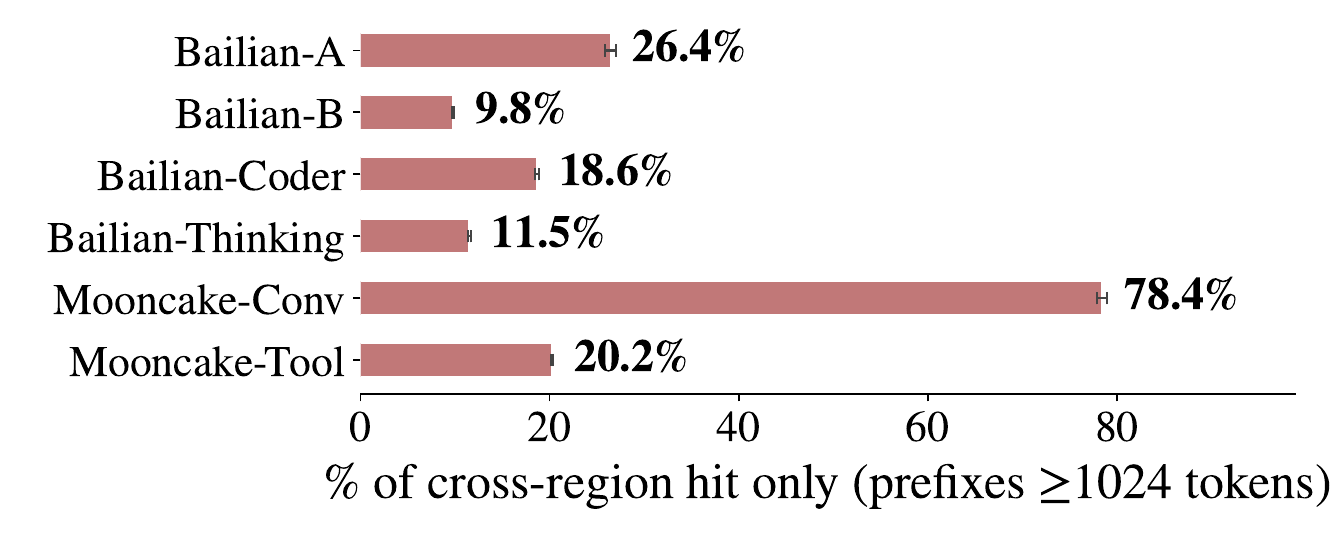}
    \caption{Cross-region only KV cache hits of ${\geq}1024$-token prefixes in production traces from Qwen-Bailian~\cite{kvcache-wild} and Mooncake~\cite{mooncake}.}
    \label{fig:motivation-traces}
    \Description{Horizontal bars show the share of reusable prefixes held only in other assigned regions for four Qwen-Bailian traces and two Mooncake traces. The shares range from about 10 to 78 percent.}
  \end{minipage}
\end{figure}

Section~\ref{sec:background:crossregion} measures the cross-region reuse opportunity on WildChat. Here we repeat the analysis on production KV cache traces from two other providers, to show the opportunity is not specific to one workload. We replay four Qwen-Bailian usage traces~\cite{kvcache-wild} with chat, coder, and reasoning traffic, and the Mooncake conversation and tool-agent traces~\cite{mooncake}. These traces carry no user locations. We assign each root session to one of 16 North-American regions in proportion to population, and follow-up turns inherit its region. We replay first-turn requests in arrival order and classify each reusable prefix of at least 1024 tokens as cold, cached in the local region, or cached only in another region, over 20 seeds. Figure~\ref{fig:motivation-traces} reports the remote-only share of hits. Across the six traces, 10 to 78\% of reusable prefixes are cached only in another region when a request arrives, and 96 to 100\% of shared-prefix reuse volume involves users from more than one region.  These additional traces support cross-region reuse opportunities under the assigned region model, rather than measure actual cross-region traffic from these providers.

